\documentclass[11pt]{article}%
\usepackage{adjustbox}
\usepackage{tabularx}
\usepackage{bm}
\usepackage{amsfonts}
\usepackage[colorlinks,citecolor=blue,urlcolor=blue,hyperfootnotes=false]%
{hyperref}
\usepackage{amssymb}
\usepackage[bottom]{footmisc}
\usepackage[hyperref]{ntheorem}
\usepackage{natbib}
\usepackage{graphicx}
\usepackage{float}
\usepackage[doublespacing,nodisplayskipstretch]{setspace}
\usepackage[top=1in, bottom=1in, left=1in, right=1in]{geometry}
\usepackage{booktabs}
\usepackage{threeparttable}
\usepackage{amsmath}
\usepackage{lscape}
\usepackage{scrextend}
\usepackage{relsize}
\usepackage{multicol}
\usepackage{chngcntr}
\usepackage{etoolbox}
\usepackage{amssymb}
\usepackage{tabularx}
\usepackage[english]{babel}
\usepackage{multirow}
\usepackage{booktabs}
\usepackage[labelfont=bf]{caption}
\usepackage{float}
\usepackage{titlesec}
\usepackage{array}
\usepackage{color}
\usepackage{afterpage}
\usepackage{framed}
\usepackage{float}
\usepackage{bigints}
\usepackage[titletoc,title, toc,page]{appendix}
\usepackage{bibunits}
\usepackage{xr}
\usepackage{hyphenat}
\usepackage{autobreak}
\usepackage{newtxtext}
\usepackage[defaultsans]{lato}%
\providecommand{\U}[1]{\protect\rule{.1in}{.1in}}
\newcolumntype{x}[1]{>{\centering \arraybackslash \hspace{0pt}}p{#1}}
\allowdisplaybreaks
\theoremstyle{plain}
\newtheorem{theorem}{Theorem}

\newtheorem{lemma}{Lemma}

\newtheorem{assumption}{Assumption}

\theorembodyfont{\normalfont}
\newtheorem{Rem}{Remark}
\deffootnote{1em}{1.6em}{\thefootnotemark \enskip}

\newcommand\independent{\protect\mathpalette{\protect\independentT}{\perp}}
\def\independentT#1#2{\mathrel{\rlap{$#1#2$}\mkern2mu{#1#2}}}

\definecolor{lightgray}{gray}{0.9}
\numberwithin{equation}{section}
\numberwithin{theorem}{section}
\numberwithin{Rem}{section}
\numberwithin{lemma}{section}
\begin{document}

\title{Covariate Distribution Balance via Propensity Scores\thanks{We thank Harold
Chiang, Cristine Pinto, Yuya Sasaki, Ping Yu, and the seminar and conference
participants at many institution for their valuable comments and suggestions.
}}
\author{Pedro H. C. Sant'Anna\thanks{Department of Economics, Vanderbilt University.
E-mail: pedro.h.santanna@vanderbilt.edu. Part of this article was written when
I was vising the Cowles Foundation at Yale University, whose hospitality is
gratefully acknowledged.}\\Vanderbilt University
\and Xiaojun Song\thanks{Department of Business Statistics and Econometrics,
Guanghua School of Management and Center for Statistical Science, Peking
University. E-mail: sxj@gsm.pku.edu.cn.}\\Peking University
\and Qi Xu\thanks{Department of Economics, Vanderbilt University. E-mail:
qi.xu.1@vanderbilt.edu.}\\Vanderbilt University}
\date{April 3, 2020}
\maketitle

\begin{abstract}
This paper proposes new estimators for the propensity score that aim to
maximize the covariate distribution balance among different treatment groups.
Heuristically, our proposed procedure attempts to estimate a propensity score
model by making the underlying covariate distribution of different treatment
groups as close to each other as possible. Our estimators are data-driven, do
not rely on tuning parameters such as bandwidths, admit an asymptotic linear
representation, and can be used to estimate different treatment effect
parameters under different identifying assumptions, including unconfoundedness
and local treatment effects. We derive the asymptotic properties of inverse
probability weighted estimators for the average, distributional, and quantile
treatment effects based on the proposed propensity score estimator and
illustrate their finite sample performance via Monte Carlo simulations and two
empirical applications.

\end{abstract}

\setlength{\abovedisplayskip}{4pt} \setlength{\belowdisplayskip}{4pt}
\setlength{\abovedisplayshortskip}{4pt} \setlength{\belowdisplayshortskip}{4pt}

\pagebreak

\section{Introduction}

Identifying and estimating the effect of a policy, treatment or intervention
on an outcome of interest is one of the main goals in applied research.
Although a randomized control trial (RCT) is the gold standard to identify
causal effects, many times its implementation is infeasible and researchers
have to rely on observational data. In such settings, the propensity score
(PS), which is defined as the probability of being treated given observed
covariates, plays a prominent role. Statistical methods using the PS include
matching, inverse probability weighting (IPW), regression, as well as
combinations thereof; for review, see, e.g., \cite{Imbens2015}.

To use these methods in practice, one has to acknowledge that the PS is
usually unknown and has to be estimated from the observed data. Given the
moderate or high dimensionality of available covariates, researchers are
usually coerced to adopt a parametric model for the PS. A popular approach is
to assume a linear logistic model, estimate the unknown parameters by maximum
likelihood (ML), check if the resulting PS estimates balance specific moments
of covariates, and in case they do not, refit the PS model including
higher-order and interaction terms and repeat the procedure until covariate
balancing is achieved, see, e.g., \cite{Rosenbaum1984} and \cite{Dehejia2002}.
On top of involving \textit{ad hoc} choices of model refinements, such model
selection procedures may result in distorted inference about the parameters of
interest, see, e.g., \cite{Leeb2005}. An additional challenge faced by PS
estimators based on ML is that the likelihood loss function does not take into
account the covariate balancing property of the PS (%
\citealp{Rosenbaum1983}%
), and, as a result, treatment effect estimators based of ML PS estimates can
be very sensitive to model misspecifications, see, e.g., \cite{Kang2007}.

In light of these practical issues, alternative estimation procedures that are
able to resemble randomization in a closer fashion have been proposed. For
instance, \cite{Graham2012}, \cite{Hainmueller2012}, \cite{Imai2014},
\cite{Zubizarreta2015}, and \cite{Zhao2018} propose alternative estimation
procedures that attempt to directly balance covariates among the treated,
untreated and, combined sample. Although such methods usually lead to
treatment effect estimators with improved finite sample properties, they only
aim to balance \textit{some} \textit{specific} functions of covariates.
However, the covariate balancing property of the PS is considerably more
powerful as it implies balance not only for some particular moments but for
\textit{all} measurable, integrable functions of the covariates. Indeed, the
balancing property of the propensity score resembles randomization: when the
data come from a randomized control trial (RCT) with perfect compliance, the
entire covariate distributions among different treatment groups are balanced
and, therefore, all measurable, integrable functions of the covariates are
indeed balanced.

In this paper, we propose an alternative framework for estimating the PS that
is arguably more suitable for causal inference, as it fully exploits the
covariate balancing property of the PS. We call the resulting PS estimator the
integrated propensity score (IPS). At a conceptual level, the IPS builds on
the observation that the covariate balancing property of the PS can be
equivalently characterized by balancing covariate distributions, namely, by an
infinite, but tractable, number of unconditional moment restrictions. Upon
such an observation, we consider Cram\'{e}r-von Mises-type distances between
these infinite balancing conditions and zero, and show that their minima are
uniquely achieved at the true PS parameters. These results, in turn, suggest
that we can estimate the unknown PS parameters within the minimum distance
framework, as in, for example, \cite{Dominguez2004} and \cite{Escanciano2006b,
Escanciano2018}. We emphasize that the IPS can be used under different
\textquotedblleft research designs\textquotedblright, including not only the
unconfounded treatment assignment setup, see, e.g., \cite{Rosenbaum1983},
\cite{Hirano2003}, \cite{Firpo2007}, and \cite{Chen2008}, but also the
\textquotedblleft local treatment effect\textquotedblright\ setup, where
selection into treatment is possibly endogenous but a binary instrumental
variable is available, see, e.g., \cite{Abadie2003}, and \cite{Frolich2013}.
In this latter case, the IPS aims to balance the covariates among the treated,
non-treated, and overall complier subpopulations.

At the practical level, one can think of the IPS as an estimation procedure
that attempts to estimate the unknown finite dimensional parameters of a PS
model by making the underlying entire covariate distribution of different treatment
groups as close to each other as possible. The IPS framework also acknowledges
that, in practice, there are different ways to compare covariate distribution
functions depending on how covariate distribution balance is measured and the
norm chosen. We explicitly consider three natural ways to characterize covariate
distribution balance: 1) using the covariates' joint cumulative distribution,
2) their joint characteristic function, or 3) exploiting the Cram\'{e}r--Wold
theorem to focus on the cumulative distribution of the one-dimensional
projections of the covariates. In terms of the norm, we focus on
Cram\'{e}r-von Mises-type distances as they can lead to smooth criteria
functions that admit a closed-form representation, allowing us to avoid using
computationally heavy numerical integration procedures. In fact, our proposed method is computationally simple and easy to use as
currently implemented in the new package \texttt{IPS} for \texttt{R},
available at \url{https://github.com/pedrohcgs/IPS}.

The proposed IPS enjoys several appealing properties. First, the IPS procedure
guarantees that the unknown PS parameters are globally identified. This is in
contrast to the traditional generalized method of moments approach based on
only finitely many balancing conditions, see, e.g., \cite{Hellerstein1999} and
\cite{Dominguez2004}. Second, even though we aim to balance an infinite number
of balancing conditions, the IPS estimator does not rely on tuning parameters.
Third, the IPS does not rely on outcome data and separates the design stage
(where one estimates the propensity score) from the analysis stage (where one
estimates different treatment effect measures). As advocated by
\cite{Rubin2007, Rubin2008}, this separation is useful as it simultaneously
mimics RCTs and avoids potential data snooping. Another direct consequence of
this clear separation is that one can use the IPS to estimate a variety of
causal effect parameters in a relatively straightforward manner. We illustrate
this flexibility by deriving the asymptotic properties of inverse probability
weighted (IPW) estimators for average, distributional and quantile treatment
effects based on the IPS, under both the unconfoundedness and the local
treatment effects setups.\medskip

\textbf{Related literature: }Our proposal builds on different branches of the
econometrics literature. For instance, this paper is related to
\cite{Shaikh2009} and \cite{SantAnna2018}, who exploit the covariate balancing
of the PS to propose specification tests for a given PS model. Here, instead
of checking if a given PS estimator balances the covariate distribution among
different treatment groups, we propose to estimate the PS unknown parameters
by maximizing the covariate balancing. The IPS estimators also build on
\cite{Dominguez2004} and \cite{Escanciano2006b, Escanciano2018}, who propose
generic estimation procedures for finite-dimensional parameters defined via an
infinite number of unconditional moment restrictions. Upon characterizing the
covariate balancing property of the PS as an infinite number of unconditional
moment restrictions, we are able to adapt their proposals to our causal
inference context.

Our proposal is also related to the growing literature on weighting-based
covariate balancing methods. Among this branch of the literature, the closest
papers to ours are \cite{Graham2012}, \cite{Imai2014}, \cite{Diaz2015} and
\cite{Fan2016a}. An important difference between our proposal and theirs is
that all these papers focus exclusively on average treatment effects under
unconfoundedness, whereas we show that one can directly use the IPS to
estimate a variety of causal parameters of interest such as average, quantile
and distributional treatment effects, not only under unconfoundedness but also
in settings with endogenous treatment. It is also worth stressing that
\cite{Graham2012} and \cite{Imai2014} propose estimating PS by balancing some
specific pre-determined moments of the covariates, and that their procedure
requires one to \emph{assume} that the propensity score parameters are
uniquely (globally) identified, see, e.g., Assumption 2.1(i) in
\cite{Graham2012}. In practice, it is hard to verify such important condition,
and when such assumption is not satisfied, inference procedures based on their
proposal will in general not be valid, see, e.g. \cite{Dominguez2004}. Our
proposed IPS procedure, on the other hand, does not suffer from this drawback
as it aims to balance the entire covariate distribution, i.e., our proposal is
based on an infinite number of balancing conditions that fully characterize
the propensity score.

In a recent working paper, \cite{Fan2016a} consider the case where the number
of balancing moments grows with the sample size at an appropriate rate.
Although this proposal bypass the identification challenge mentioned above
(see, e.g., \cite{Ai2003} and \cite{Donald2003}), to implement their proposal
one needs to carefully choose tuning parameters and select basis functions
such that the resulting balancing moments are guaranteed to be finite. Their
proposal also (implicitly) relies on covariates having compact support. Our
proposal avoids these practical complications.\medskip

\textbf{Organization of the paper: }Section \ref{sec2} introduces the
framework of balancing weights and explains the estimation problem of the IPS.
Section \ref{sec:large} presents the large sample properties of the IPS
estimator. This section also discusses how one can use the IPS to estimate and
make inference about average, distributional and quantile treatment effects
under the unconfoundedness assumption. In Section \ref{sec:endog}, we discuss
how one can use the IPS in the empirically relevant situation where treatment
adoption is endogenous and one has access to a binary instrumental variable.
Section \ref{simu} illustrates the comparative performance of the proposed
method through simulations. Section \ref{sec:application} presents two
empirical applications. Section \ref{sec:conc} concludes. Proofs, as well as
additional results, are reported in the Supplemental Appendix\footnote{The
Supplemental Appendix is available at
\url{https://pedrohcgs.github.io/files/IPS-supplementary.pdf}}.

\section{Covariate balancing via propensity score\label{sec2}}

\subsection{Background\label{sec2.1}}

Let $D$ be a binary random variable that indicates participation in the
program, i.e., $D=1$ if the individual participates in the treatment and $D=0$
otherwise. Define $Y\left(  1\right)  $ and $Y\left(  0\right)  $ as the
potential outcomes under treatment and untreated, respectively. The realized
outcome of interest is $Y=DY\left(  1\right)  +\left(  1-D\right)  Y\left(
0\right)  $, and $\mathbf{X}$\textbf{\ }is an observable $k\times1$ vector of
pre-treatment covariates.\ Denote the support of $\mathbf{X}$ by
$\mathcal{X\subset}\mathbb{R}^{k}$ and the propensity score $p\left(
\mathbf{x}\right)  =\mathbb{P}\left(  D=1|\mathbf{X}=\mathbf{x}\right)  $. For
$d\in\left\{  0,1\right\}  $, denote the distribution and quantile of the
potential outcome $Y\left(  d\right)  $ by $F_{Y\left(  d\right)  }\left(
y\right)  =\mathbb{P}\left(  Y\left(  d\right)  \leq y\right)  $, and
$q_{Y\left(  d\right)  }\left(  \tau\right)  =\inf\left\{  y:F_{Y\left(
d\right)  }\left(  y\right)  \geq\tau\right\}  $, respectively, where
$y\in\mathbb{R}$ and $\tau\in\left(  0,1\right)  $. Henceforth, assume that we
have a random sample $\left\{  \left(  Y_{i},D_{i},\mathbf{X}_{i}^{\prime
}\right)  ^{\prime}\right\}  _{i=1}^{n}$ from $\left(  Y,D,\mathbf{X}^{\prime
}\right)  ^{\prime},$ where $n\geq1$ is the sample size, and all random
variables are defined on a common probability space $\left(  \Omega
,\mathcal{A},\mathbb{P}\right)  .$ For a generic random variable $Z$, denote
$\mathbb{E}_{n}\left[  Z\right]  =n^{-1}\sum_{i=1}^{n}Z_{i}$.

The main goal in causal inference is to assess the effect of a treatment $D$
on the outcome of interest $Y$. Perhaps the most popular causal parameter of
interest is the overall average treatment effect, $ATE=\mathbb{E}\left[
Y\left(  1\right)  -Y\left(  0\right)  \right]  $. Despite its popularity, the
ATE can mask important treatment effect heterogeneity across different
subpopulations, see, e.g., \cite{Bitler2006}. Thus, in order to uncover
potential treatment effect heterogeneity, one usually focuses on different
treatment effect parameters beyond the mean. Leading examples include the
overall distributional treatment effect, $DTE\left(  y\right)  =F_{Y\left(
1\right)  }\left(  y\right)  -F_{Y\left(  0\right)  }\left(  y\right)  $, and
the overall quantile treatment effect, $QTE\left(  \tau\right)  =q_{Y\left(
1\right)  }\left(  \tau\right)  -q_{Y\left(  0\right)  }\left(  \tau\right)
$. Given that these causal parameters depend on potential outcomes that are
not jointly observed for the same individual, one cannot directly rely on the
analogy principle to identify and estimate such functionals.

A commonly used identification strategy in policy evaluation to bypass this
difficulty is to assume that selection into treatment is based on observable
characteristics, and that all individuals have a positive probability of being
in either the treatment or the untreated group --- the so-called
unconfoundedness setup, see, e.g., \cite{Rosenbaum1983}. Formally,
unconfoundedness requires the following assumption.

\begin{assumption}
\label{ignor}$(a)$ Given $\mathbf{X}$, $\left(  Y\left(  1\right)  ,Y\left(
0\right)  \right)  $ is jointly independent from $D$; and $\left(  b\right)  $
for all $\mathbf{x}\in\mathcal{X}$, $p\left(  \mathbf{x}\right)  $ is
uniformly bounded away from zero and one.
\end{assumption}

\cite{Rosenbaum1987a} shows that, under Assumption \ref{ignor}, the ATE is
identified by%
\[
ATE=\mathbb{E}\left[  \left(  \frac{D}{p\left(  \mathbf{X}\right)  }%
-\frac{\left(  1-D\right)  }{1-p\left(  \mathbf{X}\right)  }\right)  Y\right]
.
\]
Analogously, for $d\in\left\{  0,1\right\}  $, $F_{Y\left(  d\right)  }\left(
y\right)  $ is identified by
\[
F_{Y\left(  d\right)  }\left(  y\right)  =\mathbb{E}\left[  \frac{1\left\{
D=d\right\}  }{dp\left(  \mathbf{X}\right)  +(1-d)\left(  1-p\left(
\mathbf{X}\right)  \right)  }1\left\{  Y\leq y\right\}  \right]  ,
\]
with $1\left\{  \cdot\right\}  $ the indicator function, implying that
both\text{ }DTE$\left(  y\right)  $ and QTE$\left(  \tau\right)  $ can also be
written as functionals of the observed data; see, e.g., \cite{Firpo2007}, and
\cite{Chen2008}.

These identification results suggest that, if the PS were known, one could get
consistent estimators by using the sample analogue of such estimands. For
instance, one can estimate the ATE using the \cite{Hajek1971}-type estimator%
\[
\widetilde{ATE}_{n}=\mathbb{E}_{n}\left[  \left(  \varpi_{n,1}^{ps}\left(
D,\mathbf{X}\right)  -\varpi_{n,0}^{ps}\left(  D,\mathbf{X}\right)  \right)
Y\right]  ,
\]
where
\[
\varpi_{n,1}^{ps}\left(  D,\mathbf{X}\right)  =\left.  \frac{D}{p\left(
\mathbf{X}\right)  }\right/  \mathbb{E}_{n}\left[  \frac{D}{p\left(
\mathbf{X}\right)  }\right]  ,\text{ and }\varpi_{n,0}^{ps}\left(
D,\mathbf{X}\right)  =\left.  \frac{1-D}{1-p\left(  \mathbf{X}\right)
}\right/  \mathbb{E}_{n}\left[  \frac{1-D}{1-p\left(  \mathbf{X}\right)
}\right]  .
\]
Estimators for $F_{Y\left(  d\right)  }\left(  y\right)  $, $d\in\left\{
0,1\right\}  $, and DTE$\left(  y\right)  $ are formed using an analogous
strategy. For the QTE$\left(  \tau\right)  ,$ one can simply invert the
estimator of $F_{Y\left(  d\right)  }\left(  y\right)  $ to estimate
$q_{Y\left(  d\right)  }\left(  \tau\right)  $; see, e.g., \cite{Firpo2007}
and \cite{Chen2008}. Of course, estimators for other treatment effect measures
such as the difference of Theil indexes and/or Gini coefficients can also be
formed using a similar strategy, see, e.g., \cite{Firpo2016}.

In observational studies, however, the propensity score $p\left(
\mathbf{X}\right)  $ is usually unknown, and has to be estimated. Given that
$\mathbf{X}$ is usually of moderate or high dimensionality, researchers
routinely adopt a parametric approach. A popular choice among practitioners is
to use the logistic model, where%
\[
p\left(  \mathbf{X}\right)  =p\left(  \mathbf{X};\boldsymbol{\beta}%
_{0}\right)  =\frac{\exp{(\mathbf{X}^{\prime}}\boldsymbol{\beta}_{0}{)}%
}{1+\exp{(\mathbf{X}^{\prime}}\boldsymbol{\beta}_{0}{)}},
\]
with $\boldsymbol{\beta}_{0}\in\Theta\subset\mathbb{R}^{k}.$ Next, one usually
proceeds to estimate $\boldsymbol{\beta}_{0}$ within the maximum likelihood
paradigm, i.e.,%
\[
\boldsymbol{\widehat{\beta}}_{n}^{mle}=\arg\max_{\boldsymbol{\beta}\in\Theta
}\mathbb{E}_{n}\left[  D\ln\left(  p\left(  \mathbf{X};\boldsymbol{\beta
}\right)  \right)  +\left(  1-D\right)  \ln\left(  1-p\left(  \mathbf{X}%
;\boldsymbol{\beta}\right)  \right)  \right]  ,
\]
and uses the resulting PS fitted values $p\left(  \mathbf{X}%
;\boldsymbol{\widehat{\beta}}_{n}^{mle}\right)  $ to construct different
treatment effect estimators. Despite the popularity of this procedure, it has
been shown that it can lead to significant instabilities under mild PS
misspecifications, particularly when some PS estimates are relatively close to
zero or one, see e.g. \cite{Kang2007}.

In light of these challenges, alternative methods to estimate the PS have
emerged. A particularly fruitful direction is to exploit the covariate
balancing property of the PS, that is, to exploit the fact that, for all
measurable and integrable function $f\left(  \mathbf{X}\right)  $ of the
covariates $\mathbf{X}$,
\begin{equation}
\mathbb{E}\left[  \frac{D}{p\left(  \mathbf{X};\boldsymbol{\beta}_{0}\right)
}f\left(  \mathbf{X}\right)  \right]  =\mathbb{E}\left[  \frac{1-D}{1-p\left(
\mathbf{X};\boldsymbol{\beta}_{0}\right)  }f\left(  \mathbf{X}\right)
\right]  =\mathbb{E}\left[  f\left(  \mathbf{X}\right)  \right]
\label{cov.bal}%
\end{equation}
for a unique value $\boldsymbol{\beta}_{0}\in\Theta$. For example,
\cite{Imai2014} propose estimating the PS parameters $\boldsymbol{\beta}_{0}$
within the generalized method of moments framework where, for a finite vector
of user-chosen functions $f\left(  \mathbf{X}\right)  $ (e.g. $f\left(
\mathbf{X}\right)  =\mathbf{X}$),\vspace{-5pt}
\begin{equation}
\mathbb{E}\left[  \left(  \frac{D}{p\left(  \mathbf{X};\boldsymbol{\beta}%
_{0}\right)  }-\frac{1-D}{1-p\left(  \mathbf{X};\boldsymbol{\beta}_{0}\right)
}\right)  f\left(  \mathbf{X}\right)  \right]  =\mathbf{0}. \label{cbps}%
\end{equation}
\cite{Graham2012}, on the other hand, propose estimating $\boldsymbol{\beta
}_{0}$ as the solution to a globally concave programming problem such
that\vspace{-5pt}
\[
\mathbb{E}\left[  \left(  \frac{D}{p\left(  \mathbf{X};\boldsymbol{\beta}%
_{0}\right)  }-1\right)  \mathbf{X}\right]  =\mathbf{0}.
\]
Note that both procedures rely on choosing a finite number of functions
$f\left(  \mathbf{X}\right)  $, though there is little to no theoretical
guidance on how to choose such functions.

While estimators that balance low-order moments of covariates usually enjoy
more attractive finite sample properties than those based on the ML paradigm,
it is important to emphasize that the aforementioned proposals do not fully
exploit the covariate balancing property characterized in (\ref{cov.bal}).
Furthermore, as emphasized by \cite{Dominguez2004}, the global identification
condition for $\boldsymbol{\beta}_{0}$ can fail when one adopts the
generalized method of moment approach, and only attempts to balance finitely
many covariate moments.

In this paper we aim to estimate the PS parameters $\boldsymbol{\beta}_{0}$ by
taking advantage of all the information contained in (\ref{cov.bal}). Our
proposed estimators do not rely on tuning parameters such as bandwidth, do not
consult the outcome data, and can be implemented in a data-driven manner. Our
estimation procedure also guarantees that the unknown PS parameters are
globally identified.

\subsection{The integrated propensity score\label{sec:ips}}

In this section, we discuss how we operationalize our proposal. The crucial
step is to reexpress the infinite number of covariate balancing conditions
(\ref{cov.bal}) in terms of a more tractable set of moment restrictions, and
then characterize $\boldsymbol{\beta}_{0}$ as the unique minimizer of a
(population) minimum distance function. We then leverage on this
characterization, and make use of the analogy principle to suggest a natural
estimator for $\boldsymbol{\beta}_{0}$. In what follows, we present a
step-by-step description of how we achieve this.

First, note that by using the definition of conditional expectation,
(\ref{cov.bal}) can be expressed as%
\begin{equation}
\mathbb{E}\left[  \left.  \mathbf{h}\left(  D,\mathbf{X};\boldsymbol{\beta
}_{0}\right)  \right\vert \mathbf{X}\right]  =\mathbf{0}~a.s.,
\label{cov.bal2}%
\end{equation}
where $\mathbf{h}\left(  D,\mathbf{X};\boldsymbol{\beta}\right)  =\left(
h_{1}\left(  D,\mathbf{X};\boldsymbol{\beta}\right)  ,h_{0}\left(
D,\mathbf{X};\boldsymbol{\beta}\right)  \right)  ^{\prime}$, $h_{d}\left(
D,\mathbf{X};\boldsymbol{\beta}\right)  =\varpi_{d}^{ps}\left(  D,\mathbf{X}%
;\boldsymbol{\beta}\right)  -1$, $d\in\left\{  0,1\right\}  $, and
\[
\varpi_{1}^{ps}\left(  D,\mathbf{X};\boldsymbol{\beta}\right)  =\left.
\dfrac{D}{p\left(  \mathbf{X};\boldsymbol{\beta}\right)  }\right/
\mathbb{E}\left[  \dfrac{D}{p\left(  \mathbf{X};\boldsymbol{\beta}\right)
}\right]  ,\text{ }\varpi_{0}^{ps}\left(  D,\mathbf{X};\boldsymbol{\beta
}\right)  =\left.  \dfrac{1-D}{1-p\left(  \mathbf{X};\boldsymbol{\beta
}\right)  }\right/  \mathbb{E}\left[  \dfrac{1-D}{1-p\left(  \mathbf{X}%
;\boldsymbol{\beta}\right)  }\right]  .
\]
That is, one can express the covariate balancing conditions (\ref{cov.bal}) in
terms of \textit{stabilized} conditional moment restrictions.

Next, by exploiting the \textquotedblleft integrated conditional moment
approach\textquotedblright\ commonly adopted in the specification testing
literature (\citealp{Gonzalez-Manteiga2013} contains a comprehensive review),
one can express (\ref{cov.bal2}) as an infinite number of unconditional
covariate balancing restrictions. That is, by appropriately choosing a
parametric family of functions $\mathcal{W=}\left\{  w(\mathbf{X}%
;\mathbf{u}):\mathbf{u}\in\Pi\right\}  $, one can equivalently characterize
(\ref{cov.bal}) as \vspace{-15pt}
\begin{equation}
\mathbb{E}\left[  \mathbf{h}\left(  D,\mathbf{X};\boldsymbol{\beta}%
_{0}\right)  w(\mathbf{X};\mathbf{u})\right]  =\mathbf{0}~a.e~in~\mathbf{u}%
\in\Pi, \label{cov.bal3}%
\end{equation}
see, e.g., Lemma 1 of \cite{Escanciano2006a} for primitive conditions on the
family $\mathcal{W}$ such that the equivalence between (\ref{cov.bal2}) and
(\ref{cov.bal3}) holds. Choices of weight $w$ satisfying this equivalence
include $(a)$ $w(\mathbf{X};\mathbf{u})=1\left\{  \mathbf{X}\leq
\mathbf{u}\right\}  $, where $\mathbf{u}\in\left[  -\infty,\infty\right]
^{k}$, $1\left\{  A\right\}  $ denotes the indicator function of the event $A$
and $\mathbf{X}\leq\mathbf{u}$ is understood coordinate-wise (see, e.g.,
\citealp{Stute1997} and \citealp{Dominguez2004, Dominguez2015}), $(b)$
$w(\mathbf{X};\mathbf{u})=\exp(i\mathbf{u}^{\prime}\Phi\left(  \mathbf{X}%
\right)  \mathbf{)}$, where $\mathbf{u}$ $\in$ $\mathbb{R}^{k}$, $\Phi\left(
\cdot\right)  $ is a vector of bounded one-to-one maps from $\mathbb{R}^{k}$
to $\mathbb{R}^{k}$ and $i=\sqrt{-1}$ is the imaginary unit (see, e.g.,
\citealp{Bierens1982} and \citealp{Escanciano2018}), and $\left(  c\right)  $
$w(\mathbf{X};\mathbf{u})=1\left\{  \boldsymbol{\gamma}^{\prime}\mathbf{X}\leq
u\right\}  $,$~$where~$\mathbf{u=}\left(  \boldsymbol{\gamma},u\right)
\in\mathbb{S}_{k}\times\left[  -\infty,\infty\right]  $, $\mathbb{S}%
_{k}=\left\{  \boldsymbol{\gamma}\in\mathbb{R}^{k}:\left\Vert
\boldsymbol{\gamma}\right\Vert =1\right\}  $, and $\left\Vert
\boldsymbol{\gamma}\right\Vert $ is the Euclidean norm of real-valued vector
$\boldsymbol{\gamma}$ (see, e.g., \citealp{Escanciano2006b}). We call
(\ref{cov.bal3}) the \textquotedblleft integrated covariate balancing
condition\textquotedblright\ because it uses the integrated (cumulative)
measure of covariate balancing.

Finally, let\vspace{-5pt}
\begin{equation}
Q_{w}\left(  \boldsymbol{\beta}\right)  =\int_{\Pi}\left\Vert \mathbf{H}%
_{w}(\boldsymbol{\beta},\mathbf{u})\right\Vert ^{2}\,\Psi(d\mathbf{u}%
),\quad\boldsymbol{\beta}\in\Theta\subset\mathbb{R}^{k}, \label{min.dist}%
\end{equation}
where $\mathbf{H}_{w}(\boldsymbol{\beta},\mathbf{u})=\mathbb{E}\left[
\mathbf{h}\left(  D,\mathbf{X};\boldsymbol{\beta}\right)  w(\mathbf{X}%
;\mathbf{u})\right]  $, $\left\Vert A\right\Vert ^{2}=A^{c}A$, $A^{c}$ denotes
the conjugate transpose of the column vector $A$, and $\Psi(\mathbf{u})$ is an
integrating probability measure that is absolutely continuous with respect to
a dominating measure on $\Pi$.

With these results in hand, in the following lemma we show that \vspace{-5pt}
\begin{equation}
\boldsymbol{\beta}_{0}=\arg\min_{\boldsymbol{\beta}\in\Theta}Q_{w}\left(
\boldsymbol{\beta}\right)  , \label{obj}%
\end{equation}
and $\boldsymbol{\beta}_{0}$ is the unique value such that the covariate
balancing condition (\ref{cov.bal}) is satisfied.

\begin{lemma}
\label{lemma1}Let $\Theta\subset\mathbb{R}^{k}$ be the parameter space, and
assume that (\ref{cov.bal}) is satisfied for a unique $\boldsymbol{\beta}%
_{0}\in\Theta$. Then $Q_{w}(\boldsymbol{\beta})\geq0$, $\forall
\boldsymbol{\beta}\in\Theta$, and $Q_{w}(\boldsymbol{\beta}_{0})=0$ if and
only if the covariate balancing condition (\ref{cov.bal}) holds.
\end{lemma}

Lemma \ref{lemma1} is a global identification result that characterizes
$\boldsymbol{\beta}_{0}$ as the unique minimizer of a population minimum
distance function, $Q_{w}(\boldsymbol{\beta})$. That is, from Lemma
\ref{lemma1} we have that $\boldsymbol{\beta}_{0}$ is the unique PS parameter
that minimizes the imbalances of all measurable and integrable functions
$f\left(  \mathbf{X}\right)  $ between the treated, untreated and the combined
group. Here, it is worth mentioning that neither \cite{Graham2012} nor
\cite{Imai2014} covariate balancing approach guarantee global identification
of the propensity score parameters. Instead, they directly \emph{assume} that
the vector of user-selected balancing conditions uniquely identify the
propensity score parameters; see, e.g., Assumption 2.1 (i) of
\cite{Graham2012}. In practice, however, it is hard if not impossible to
verify if such condition indeed holds. In cases it does not hold, inference
procedures that rely on their proposed propensity score estimator, in general,
will not be valid; see, e.g., \cite{Dominguez2004}. Lemma \ref{lemma1} shows
that our propose IPS procedure completely avoids this important drawback.

Another important implication of Lemma \ref{lemma1} is that it suggests a
natural estimator for $\boldsymbol{\beta}_{0}$ based on the sample analogue of
(\ref{obj}), namely,\vspace{-15pt}
\begin{equation}
\widehat{\boldsymbol{\beta}}_{n,w}^{ips}=\arg\min_{\boldsymbol{\beta}\in
\Theta}Q_{n,w}(\boldsymbol{\beta}), \label{ips}%
\end{equation}
where $Q_{n,w}(\boldsymbol{\beta})=\int_{\Pi}\left\Vert \mathbf{H}%
_{n,w}(\boldsymbol{\beta},\mathbf{u})\right\Vert ^{2}\,\Psi_{n}(d\mathbf{u})$,
$\Psi_{n}$ is a uniformly consistent estimator of $\Psi$, $\mathbf{H}%
_{n,w}(\boldsymbol{\beta},\mathbf{u})=\mathbb{E}_{n}\left[  \mathbf{h}%
_{n}\left(  D,\mathbf{X};\boldsymbol{\beta}\right)  w(\mathbf{X}%
;\mathbf{u})\right]  $, with $\mathbf{h}_{n}\left(  D,\mathbf{X}%
;\boldsymbol{\beta}\right)  =\left(  h_{n,1}\left(  D,\mathbf{X}%
;\boldsymbol{\beta}\right)  ,h_{n,0}\left(  D,\mathbf{X};\right.  \right.  $
$\left.  \left.  \boldsymbol{\beta}\right)  \right)  ^{\prime}$,
$h_{n,d}\left(  D,\mathbf{X};\boldsymbol{\beta}\right)  =\varpi_{n,d}%
^{ps}\left(  D,\mathbf{X};\boldsymbol{\beta}\right)  -1$, $d\in\left\{
0,1\right\}  $, and%
\begin{align}
\varpi_{n,1}^{ps}\left(  D,\mathbf{X};\boldsymbol{\beta}\right)   &  =\left.
\dfrac{D}{p\left(  \mathbf{X};\boldsymbol{\beta}\right)  }\right/
\mathbb{E}_{n}\left[  \dfrac{D}{p\left(  \mathbf{X};\boldsymbol{\beta}\right)
}\right]  ,\label{w1}\\
\varpi_{n,0}^{ps}\left(  D,\mathbf{X};\boldsymbol{\beta}\right)   &  =\left.
\dfrac{1-D}{1-p\left(  \mathbf{X};\boldsymbol{\beta}\right)  }\right/
\mathbb{E}_{n}\left[  \dfrac{1-D}{1-p\left(  \mathbf{X};\boldsymbol{\beta
}\right)  }\right]  . \label{w0}%
\end{align}
We call $\widehat{\boldsymbol{\beta}}_{n,w}^{ips}$ the integrated propensity
score estimator of $\boldsymbol{\beta}_{0}$ because it is based on the
integrated covariate balancing conditions (\ref{cov.bal3}).

From (\ref{ips}), one can conclude that different PS estimators that fully
exploit the covariate balancing property (\ref{cov.bal}) can be constructed by
choosing different $w$ and $\Psi_{n}$. In this article, we focus on three
different combinations that are intuitive, computationally simple, and that
perform well in practice:\vspace{-15pt}

\begin{enumerate}
\item[$\left(  i\right)  $] $w(\mathbf{X};\mathbf{u})=1\left\{  \mathbf{X}%
\leq\mathbf{u}\right\}  $ and $\Psi_{n}(\mathbf{u})=F_{n,\boldsymbol{X}%
}(\mathbf{u})\equiv n^{-1}\sum_{i=1}^{n}1\left\{  \mathbf{X}_{i}\leq
\mathbf{u}\right\}  $, leading to the IPS estimator \vspace{-15pt}
\begin{equation}
\widehat{\boldsymbol{\beta}}_{n,\text{ind}}^{ips}=\arg\min_{\boldsymbol{\beta
}\in\Theta}\int_{\left[  -\infty,\infty\right]  ^{k}}\left\Vert \mathbb{E}%
_{n}\left[  \mathbf{h}_{n}\left(  D,\mathbf{X};\boldsymbol{\beta}\right)
1\left\{  \mathbf{X}\leq\mathbf{u}\right\}  \right]  \right\Vert ^{2}%
\,F_{n,X}(d\mathbf{u}); \label{ips-ind}%
\end{equation}

\item[$\left(  ii\right)  $] $w(\mathbf{X};\mathbf{u})=1\left\{
\boldsymbol{\gamma}^{\prime}\mathbf{X}\leq u\right\}  $ with $\Psi
_{n}(\mathbf{u})$ the product measure of \ $F_{n,\boldsymbol{\gamma}}\left(
u\right)  \equiv$ $n^{-1}\sum_{i=1}^{n}$ $1\left\{  \boldsymbol{\gamma
}^{\prime}\mathbf{X}_{i}\leq u\right\}  $ and the uniform distribution on
$\mathbb{S}_{k},$ leading to the IPS estimator%
\begin{equation}
\widehat{\boldsymbol{\beta}}_{n,\text{proj}}^{ips}=\arg\min_{\boldsymbol{\beta
}\in\Theta}\int_{\left[  -\infty,\infty\right]  \times\mathbb{S}_{k}%
}\left\Vert \mathbb{E}_{n}\left[  \mathbf{h}_{n}\left(  D,\mathbf{X}%
;\boldsymbol{\beta}\right)  1\left\{  \boldsymbol{\gamma}^{\prime}%
\mathbf{X}\leq u\right\}  \right]  \right\Vert ^{2}\,F_{n,\boldsymbol{\gamma}%
}(du)d\boldsymbol{\gamma~}; \label{ips-proj}%
\end{equation}

\item[$\left(  iii\right)  $] $w(\mathbf{X};\mathbf{u})=\exp(i\mathbf{u}%
^{\prime}\Phi\left(  \mathbf{X}\right)  \mathbf{)}$ with $\Psi_{n}%
(\mathbf{u})\equiv\Psi(\mathbf{u})$, the CDF of $k$-variate standard normal
distribution, $\Phi\left(  \mathbf{X}\right)  =\left(  \Phi\left(
\widetilde{X}_{1}\right)  ,\dots,\Phi\left(  \widetilde{X}_{k}\right)
\right)  ^{\prime},$ $\widetilde{X_{p}}$ the studentized $X_{p}$, and $\Phi$
the univariate CDF of the standard normal distribution, leading to the IPS
estimator%
\begin{equation}
\widehat{\boldsymbol{\beta}}_{n,\text{exp}}^{ips}=\arg\min_{\boldsymbol{\beta
}\in\Theta}\int_{\mathbb{R}^{k}}\left\Vert \mathbb{E}_{n}\left[
\mathbf{h}_{n}\left(  D,\mathbf{X};\boldsymbol{\beta}\right)  \exp
(i\mathbf{u}^{\prime}\Phi\left(  \mathbf{X}\right)  \mathbf{)}\right]
\right\Vert ^{2}\,\frac{\exp\left(  -\frac{1}{2}\mathbf{u}^{\prime}%
\mathbf{u}\right)  }{\left(  2\pi\right)  ^{k/2}}d\mathbf{u.} \label{ips-exp}%
\end{equation}

\end{enumerate}

The estimators (\ref{ips-ind})-(\ref{ips-exp}) build on \cite{Dominguez2004}
and \cite{Escanciano2006b, Escanciano2018}, respectively. Despite the apparent
differences, they all aim to minimize covariate distribution imbalances:
(\ref{ips-ind}) aims to directly minimize imbalances of the joint distribution
of covariates; (\ref{ips-proj}) exploits the Cram\'{e}r-Wold theorem and
focuses on minimizing imbalances of the distribution of all one-dimensional
projections of covariates; and (\ref{ips-exp}) focuses on minimizing
imbalances of the (transformed) covariates' joint characteristic function.
From the Cram\'{e}r-Wold theorem and the fact that the characteristic function
completely defines the distribution function (and vice-versa), (\ref{ips-ind}%
)-(\ref{ips-exp}) are indeed intrinsically related. Furthermore, we emphasize
that our estimators are data-driven, and neither $w$ nor $\Psi_{n}$ plays the
role of a bandwidth as they do not affect the convergence rate of the IPS estimator.

From the computational perspective, (\ref{ips-ind})-(\ref{ips-exp}) are easy
to estimate because they do not involve matrix inversion nor nonparametric
estimation. In the supplemental Appendix \ref{compstats}, we show that the
objective functions in (\ref{ips-ind})-(\ref{ips-exp}) can be written in
closed form, which, in turn, implies a more straightforward implementation. In
practice, the IPS is easy to use as it is already implemented in the new
package \texttt{IPS} for \texttt{R}, available at \url{https://github.com/pedrohcgs/IPS}.

\begin{Rem}
It is important to stress that the covariate balancing property (\ref{cov.bal}%
) follows directly from the definition of the PS and does not depend on the
unconfoundedness assumption \ref{ignor}. Thus, one can use our proposed IPS
estimators even in contexts where Assumption \ref{ignor} does not hold,
though, in such cases, the resulting (second step) estimators may be only
descriptive, see, e.g., \cite{Dinardo1996}, and \cite{Kline2011}. In addition,
as we discuss in Section \ref{sec:endog}, the same principle can be used to
balance the covariate distributions among the treated and non-treated complier subpopulations.
\end{Rem}

\begin{Rem}
\label{rem-3bal}It is interesting to compare (\ref{cbps}) with (\ref{cov.bal3}%
) beyond the fact that (\ref{cov.bal3}) is based on infinitely many balancing
conditions whereas (\ref{cbps}) is not. First, note that (\ref{cov.bal3}) is
based on normalized (or stabilized) weights whereas (\ref{cbps}) is not. We
prefer to use stabilized weights as treatment effect estimators based on them
usually have improved finite sample properties (see, e.g.,
\citealp{Millimet2009} and \citealp{Busso2014}). Second, note that
(\ref{cov.bal3}) implies a three-way balance (treated, untreated and combined
groups), whereas (\ref{cbps}) only imposes a two-way balance (treated and
untreated). We note that (\ref{cbps}) can lead to relatively smaller/larger PS
estimates as a \textquotedblleft close to zero\textquotedblright\ PS estimate
in the treated group can be offset by a \textquotedblleft close to
one\textquotedblright\ PS estimate in the untreated group. By using
(\ref{cov.bal3}), such a potential drawback is avoided.
\end{Rem}

\section{Large sample properties\label{sec:large}}

In this section, we first derive the asymptotic properties of the IPS
estimators, namely the consistency, asymptotic linear representation, and
asymptotic normality of $\widehat{\boldsymbol{\beta}}_{n,w}^{ips}$ . We then
discuss how one can build on these results to conduct asymptotically valid
inference for overall average, distributional and quantile treatment effects,
using inverse probability weighted estimators. Although our proposal can also
be used to estimate other treatment effects of interest such as those
discussed in \cite{Firpo2016}, we omit such a discussion for the sake of brevity.

\subsection{Asymptotic theory for IPS estimator}

Here we derive the asymptotic properties of the IPS estimator. Let the score
of $\mathbf{H}_{w}(\boldsymbol{\beta},\mathbf{u})$ be defined as
$\mathbf{\dot{H}}_{w}(\boldsymbol{\beta},\mathbf{u})=\left(  \mathbf{\dot{H}%
}_{1,w}^{^{\prime}}(\boldsymbol{\beta},\mathbf{u}),\mathbf{\dot{H}}%
_{0,w}^{^{\prime}}(\boldsymbol{\beta},\mathbf{u})\right)  ^{\prime},$ a
$2\times k$ matrix, where, for $d\in\left\{  0,1\right\}  ,$ $\mathbf{\dot{H}%
}_{d,w}(\boldsymbol{\beta},\mathbf{u})=\mathbb{E}\left[  \mathbf{\dot{h}}%
_{d}\left(  D,\mathbf{X};\boldsymbol{\beta}\right)  w(\mathbf{X}%
;\mathbf{u})\right]  $, with $\mathbf{\dot{h}}_{1}$ and $\mathbf{\dot{h}}_{0}$
being the $1\times k$ vectors defined as%
\begin{align*}
\mathbf{\dot{h}}_{1}\left(  D,\mathbf{X};\boldsymbol{\beta}\right)   &
=-\frac{\varpi_{1}^{ps}\left(  D,\mathbf{X};\boldsymbol{\beta}\right)
}{p\left(  \mathbf{X};\boldsymbol{\beta}\right)  }\dot{p}\left(
\mathbf{X};\boldsymbol{\beta}\right)  ^{\prime}+\varpi_{1}^{ps}\left(
D,\mathbf{X};\boldsymbol{\beta}\right)  \cdot\mathbb{E}\left[  \frac
{\varpi_{1}^{ps}\left(  D,\mathbf{X};\boldsymbol{\beta}\right)  }{p\left(
\mathbf{X};\boldsymbol{\beta}\right)  }\dot{p}\left(  \mathbf{X}%
;\boldsymbol{\beta}\right)  ^{\prime}\right]  ,\\
\mathbf{\dot{h}}_{0}\left(  D,\mathbf{X};\boldsymbol{\beta}\right)   &
=\frac{\varpi_{0}^{ps}\left(  D,\mathbf{X};\boldsymbol{\beta}\right)
}{1-p\left(  \mathbf{X};\boldsymbol{\beta}\right)  }\dot{p}\left(
\mathbf{X};\boldsymbol{\beta}\right)  ^{\prime}-\varpi_{0}^{ps}\left(
D,\mathbf{X};\boldsymbol{\beta}\right)  \cdot\mathbb{E}\left[  \frac
{\varpi_{0}^{ps}\left(  D,\mathbf{X};\boldsymbol{\beta}\right)  }{1-p\left(
\mathbf{X};\boldsymbol{\beta}\right)  }\dot{p}\left(  \mathbf{X}%
;\boldsymbol{\beta}\right)  ^{\prime}\right]  ,
\end{align*}
and $\dot{p}\left(  \mathbf{\cdot;}\boldsymbol{\beta}\right)  =\left.  \left.
\partial p\left(  \mathbf{\cdot;}\boldsymbol{b}\right)  \right/
\partial\boldsymbol{b}\right\vert _{\boldsymbol{b=\beta}},$ the $k\times1$
vector of scores of the PS model $p\left(  \cdot,\boldsymbol{\beta}\right)  $.
We make the following set of assumptions.

\begin{assumption}
\label{ass.pscore}$\left(  i\right)  p\left(  \mathbf{x}\right)  =p\left(
\mathbf{x};\boldsymbol{\beta}_{0}\right)  $, where $\boldsymbol{\beta}_{0}$ is
an interior point of a compact set $\Theta\subset\mathbb{R}^{k};$ $\left(
ii\right)  $ for some $\delta>0$, $\delta\leq p\left(  \mathbf{x;}%
\boldsymbol{\beta}\right)  \leq1-\delta$ for all $\mathbf{x}\in\mathcal{X}$,
$\boldsymbol{\beta}\in int\left(  \Theta\right)  $; $\left(  iii\right)  $
with probability one, $p\left(  \mathbf{X;}\boldsymbol{\beta}\right)  $ is
continuous at each $\boldsymbol{\beta}\in$ $\Theta$; $\left(  iv\right)  $
with probability one, $p\left(  \mathbf{X;}\boldsymbol{\beta}\right)  $ is
continuously differentiable in a neighborhood of $\boldsymbol{\beta}_{0}%
$,$\Theta_{0}$ $\subset\Theta$ $;$ $\left(  v\right)  $ for $d\in\left\{
0,1\right\}  $
\[
\mathbb{E}\left[  \sup_{\boldsymbol{\beta}\in\Theta_{0}}\left\Vert \left(
\frac{\varpi_{d}^{ps}\left(  D,\mathbf{X};\boldsymbol{\beta}\right)  }{d\cdot
p\left(  \mathbf{X};\boldsymbol{\beta}\right)  +\left(  1-d\right)
\cdot\left(  1-p\left(  \mathbf{X};\boldsymbol{\beta}\right)  \right)
}\right)  \cdot\dot{p}\left(  \mathbf{X};\boldsymbol{\beta}\right)
\right\Vert \right]  <\infty.
\]

\end{assumption}

\begin{assumption}
\label{ass.w} The family of weighting functions and integrating probability
measures satisfy one of the following:

$\left(  i\right)  \mathcal{W}_{ind}\mathcal{\equiv}\left\{  \mathbf{x}%
\in\mathcal{X}\mapsto1\left\{  \mathbf{x}\leq\mathbf{u}\right\}
:\mathbf{u}\in\left[  -\infty,\infty\right]  ^{k}\right\}  $, $\Psi
_{n}(\mathbf{u})=F_{n,\mathbf{X}}(\mathbf{u})$, and $\Psi(\mathbf{u}%
)=F_{\mathbf{X}}(\mathbf{u})$, where $F_{n,\mathbf{X}}(\mathbf{u})\equiv
n^{-1}\sum_{i=1}^{n}1\left\{  \mathbf{X}_{i}\leq\mathbf{u}\right\}  $, and
$F_{\mathbf{X}}(\mathbf{u})\equiv\mathbb{E}\left[  1\left\{  \mathbf{X}%
\leq\mathbf{u}\right\}  \right]  ;$

$\left(  ii\right)  \mathcal{W}_{\text{proj}}\mathcal{\equiv}\left\{
\mathbf{x}\in\mathcal{X}\mapsto1\left\{  \boldsymbol{\gamma}^{\prime
}\mathbf{x}\leq u\right\}  :\left(  \boldsymbol{\gamma},u\right)
\in\mathbb{S}_{k}\times\left[  -\infty,\infty\right]  \right\}  $, $\Psi
_{n}\left(  \mathbf{u}\right)  =F_{n,\boldsymbol{\gamma}}\left(  u\right)
\times \Upsilon$, and $\Psi\left(  \mathbf{u}\right)  =F_{\boldsymbol{\gamma}%
}\left(  u\right)  \times \Upsilon$, where $\mathbb{S}_{k}\equiv\left\{
\boldsymbol{\gamma}\in\mathbb{R}^{k}:\left\Vert \boldsymbol{\gamma}\right\Vert
=1\right\}  ,$ $F_{n,\boldsymbol{\gamma}}\left(  u\right)  \equiv n^{-1}%
\sum_{i=1}^{n}1\left\{  \boldsymbol{\gamma}^{\prime}\mathbf{X}_{i}\leq
u\right\}  $, $F_{\boldsymbol{\gamma}}\left(  u\right)  \equiv\mathbb{E}%
\left[  1\left\{  \boldsymbol{\gamma}^{\prime}\mathbf{X}\leq u\right\}
\right]  $ and $\Upsilon$ is the uniform distribution on $\mathbb{S}_{k};$

$\left(  iii\right)  \mathcal{W}_{\text{exp}}\mathcal{\equiv}\left\{
\mathbf{x}\in\mathcal{X}\mapsto\exp(i\mathbf{u}^{\prime}\Phi\left(
\mathbf{x}\right)  \mathbf{)}:\mathbf{u}\in\Pi\right\}  $, and $\Psi
_{n}(\mathbf{u})=\Psi(\mathbf{u})$, where $\Pi$ is any compact, convex subset
$\mathbb{R}^{k}$ with a non-empty interior, and $\Psi(\mathbf{u})$ is the CDF
of $k$-variate standard normal distribution.
\end{assumption}

Assumption \ref{ass.pscore} is standard in the literature, see, e.g., Theorems
2.6 and 3.4 of \cite{Newey1994c}, Example 5.40 of \cite{VanderVaart1998}, and
\cite{Graham2012}. Assumption \ref{ass.pscore}$(i)$ states that the true PS is
known up to finite dimensional parameters $\boldsymbol{\beta}_{0}$, that is,
we are in a parametric setup. Assumption \ref{ass.pscore}$\left(  ii\right)  $
imposes that the parametric PS is bounded from above and from below. This
assumption can be relaxed by assuming that $\left(  D/p\left(  \mathbf{X}%
;\boldsymbol{\beta}\right)  ,\left(  1-D\right)  /\left(  1-p\left(
\mathbf{X};\boldsymbol{\beta}\right)  \right)  \right)  ^{\prime}%
\leq\mathbf{b}\left(  \mathbf{X}\right)  $ such that $\mathbb{E}\left[
\left\Vert \mathbf{b}\left(  \mathbf{X}\right)  \right\Vert ^{2}\right]
<\infty$. Assumptions \ref{ass.pscore}$(iii)$-$\left(  iv\right)  $ impose
additional smoothness conditions on the PS, whereas Assumption
\ref{ass.pscore}$(v)$ (together with Assumption \ref{ass.w}) implies that, in
a small neighborhood of $\boldsymbol{\beta}_{0}$ and for all $u\in\Pi$, the
score $\mathbf{\dot{H}}_{w}(\boldsymbol{\beta},\mathbf{u})$ is uniformly
bounded by an integrable function.

Assumption \ref{ass.w} restricts our attention to the IPS estimators
(\ref{ips-ind})-(\ref{ips-exp}). As mentioned before, we focus on such
estimators because of their computational simplicity and transparency.
Nonetheless, other types of IPS estimators can also be formed, provided that
the weighting function $w$ and integrating measure $\Psi_{n}$ satisfy some
high-level regularity conditions.

The next theorem characterizes the asymptotic properties of the IPS estimators
$\widehat{\boldsymbol{\beta}}_{n,w}^{ips}$. Define the $k\times k$ matrix
\vspace{-5pt}
\[
C_{w,\Psi}=\int_{\Pi}\left(  \mathbf{\dot{H}}_{w}(\boldsymbol{\beta}%
_{0},\mathbf{u})^{c}~\mathbf{\dot{H}}_{w}(\boldsymbol{\beta}_{0}%
,\mathbf{u})+\mathbf{\dot{H}}_{w}(\boldsymbol{\beta}_{0},\mathbf{u})^{\prime
}\left(  \mathbf{\dot{H}}_{w}(\boldsymbol{\beta}_{0},\mathbf{u})^{\prime
}\right)  ^{c}\right)  \Psi(d\mathbf{u}),
\]
and the $k\times1$ vector \vspace{-15pt}
\begin{multline*}
l_{w,\Psi}\left(  D,\mathbf{X};\boldsymbol{\beta}_{0}\right)  =-C_{w,\Psi
}^{-1}\cdot\int_{\Pi}\left(  \mathbf{\dot{H}}_{w}(\boldsymbol{\beta}%
_{0},\mathbf{u})^{c}~w(\mathbf{X};\mathbf{u})+\mathbf{\dot{H}}_{w}%
(\boldsymbol{\beta}_{0},\mathbf{u})^{\prime}w(\mathbf{X};\mathbf{u}%
)^{c}\right)  \Psi(d\mathbf{u}) \cdot\mathbf{h}\left(  D,\mathbf{X}%
;\boldsymbol{\beta}_{0}\right)  .
\end{multline*}

\begin{theorem}
\label{th.ips}Under Assumptions \ref{ass.pscore} - \ref{ass.w}, as
$n\rightarrow\infty$,
\[
\widehat{\boldsymbol{\beta}}_{n,w}^{ips}-\boldsymbol{\beta}_{0}=o_{p}\left(
1\right)  .
\]
Furthermore, provided that the matrix $C_{w,\Psi}$ is positive definite,
\begin{equation}
\sqrt{n}\left(  \widehat{\boldsymbol{\beta}}_{n,w}^{ips}-\boldsymbol{\beta
}_{0}\right)  =\frac{1}{\sqrt{n}}\sum_{i=1}^{n}l_{w,\Psi}\left(
D_{i},\mathbf{X}_{i};\boldsymbol{\beta}_{0}\right)  +o_{p}\left(  1\right)  ,
\label{lin.rep}%
\end{equation}
and\vspace{-15pt}
\[
\sqrt{n}\left(  \widehat{\boldsymbol{\beta}}_{n,w}^{ips}-\boldsymbol{\beta
}_{0}\right)  \overset{d}{\rightarrow}N\left(  0,\Omega_{w,\Psi}^{ips}\right)
\text{,}%
\]
where $\Omega_{w,\Psi}^{ips}\equiv\mathbb{E}\left[  l_{w,\Psi}\left(
D,\mathbf{X};\boldsymbol{\beta}_{0}\right)  l_{w,\Psi}\left(  D,\mathbf{X}%
;\boldsymbol{\beta}_{0}\right)  ^{\prime}\right]  .$
\end{theorem}

From Theorem \ref{th.ips}, we conclude that the proposed IPS estimator is
consistent, admits an asymptotic linear representation with influence function
$l_{w,\Psi}\left(  D,\mathbf{X};\boldsymbol{\beta}_{0}\right)  $, and
converges to a normal distribution. The asymptotic linear representation
(\ref{lin.rep}) plays a major role in establishing the asymptotic properties
of causal parameters such as average, distributional, and quantile treatment
effects; see Section \ref{causal}.

\begin{Rem}
Although the results in Theorem \ref{th.ips} focus on the case where the
propensity score is correctly specified, it is not difficult to show that the
IPS estimators are still consistent when the model is locally misspecified,
i.e., when $\mathbb{E}\left[  \left.  \mathbf{h}\left(  D,\mathbf{X}%
;\boldsymbol{\beta}_{0}\right)  \right\vert \mathbf{X}\right]  =n^{-1/2}%
\cdot\mathbf{s}\left(  \mathbf{X}\right)  $ a.s., for some integrable function
$\mathbf{s}\left(  \mathbf{X}\right)  $. In this case, $\sqrt{n}\left(
\widehat{\boldsymbol{\beta}}_{n,w}^{ips}-\boldsymbol{\beta}_{0}\right)  $
would still be asymptotically normal, with a mean given by
\[
-C_{w,\Psi}^{-1}\cdot\int_{\Pi}\left(  \mathbf{\dot{H}}_{w}(\boldsymbol{\beta
}_{0},\mathbf{u})^{c}~\mathbf{S}_{w}\left(  \mathbf{u}\right)  +\mathbf{\dot
{H}}_{w}^{^{\prime}}(\boldsymbol{\beta}_{0},\mathbf{u})\left(  \mathbf{S}%
_{w}\left(  \mathbf{u}\right)  ^{\prime}\right)  ^{c}\right)  \Psi
(d\mathbf{u}),
\]
where $\mathbf{S}_{w}\left(  \mathbf{u}\right)  =\mathbb{E}\left[
\mathbf{s}\left(  \mathbf{X}\right)  w(\mathbf{X};\mathbf{u})\right]  $, and
variance given by $\Omega_{w,\Psi}^{ips}$; see, e.g., Remark 1 in
\cite{Escanciano2006b}, and Propositions 3 and 4 in \cite{Dominguez2015}.
Based on these results, it is straightforward to compute the local bias of IPW
estimators for different causal parameters. We omit such derivations for the
sake of brevity.
\end{Rem}

\subsection{Estimating treatment effects under unconfoundedness\label{causal}}

In this section, we illustrate how one can estimate and make asymptotically
valid inference about average, distributional, and quantile treatment effects
under the unconfoundedness assumption \ref{ignor} using IPW estimators based
on the IPS estimator $\widehat{\boldsymbol{\beta}}_{n,w}^{ips}.$

Based on the discussion in Section \ref{sec2.1}, the IPW estimators for ATE,
DTE and QTE are respectively:%
\begin{align}
\widehat{ATE}_{n}^{ips}  &  =\mathbb{E}_{n}\left[  \left(  \varpi_{n,1}%
^{ps}\left(  D,\mathbf{X};\widehat{\boldsymbol{\beta}}_{n,w}^{ips}\right)
-\varpi_{n,0}^{ps}\left(  D,\mathbf{X};\widehat{\boldsymbol{\beta}}%
_{n,w}^{ips}\right)  \right)  Y\right]  ,\label{ate.hat}\\
\widehat{DTE}_{n}^{ips}\left(  y\right)   &  =\mathbb{E}_{n}\left[  \left(
\varpi_{n,1}^{ps}\left(  D,\mathbf{X};\widehat{\boldsymbol{\beta}}_{n,w}%
^{ips}\right)  -\varpi_{n,0}^{ps}\left(  D,\mathbf{X}%
;\widehat{\boldsymbol{\beta}}_{n,w}^{ips}\right)  \right)  1\left\{  Y\leq
y\right\}  \right]  ,\label{dte.hat}\\
\widehat{QTE}_{n}^{ips}\left(  \tau\right)   &  =\widehat{q}_{n,Y\left(
1\right)  }^{ips}\left(  \tau\right)  -\widehat{q}_{n,Y\left(  0\right)
}^{ips}\left(  \tau\right)  , \label{qte.hat}%
\end{align}
where, for $d\in\left\{  0,1\right\}  $,%
\[
\widehat{q}_{n,Y\left(  d\right)  }^{ips}=\arg\min_{q\in\mathbb{R}}%
\mathbb{E}_{n}\left[  \varpi_{n,d}^{ps}\left(  D,\mathbf{X}%
;\widehat{\boldsymbol{\beta}}_{n,w}^{ips}\right)  \cdot\rho_{\tau}\left(
Y-q\right)  \right]  ,
\]
with $\rho_{\tau}\left(  a\right)  =a\cdot\left(  \tau-1\left\{
a\leq0\right\}  \right)  $ the check function as in \cite{Koenker1978}, and
the weights $\varpi_{n,1}^{ps}$ and $\varpi_{n,0}^{ps}$ are as in
(\ref{w1})-(\ref{w0}).

To derive the asymptotic properties of (\ref{ate.hat})-(\ref{qte.hat}), we
need to make an additional assumption about the underlying distributions of
the potential outcomes $Y\left(  1\right)  $ and $Y\left(  0\right)  $.

\begin{assumption}
\label{regularity}For $d\in\left\{  0,1\right\}  $, $\left(  i\right)
\mathbb{E}\left[  Y\left(  d\right)  ^{2}\right]  <M$ for some $0<M<\infty$,
$\left(  ii\right)  $
\[
\mathbb{E}\left[  \sup_{\beta\in\Theta_{0}}\left\Vert \dfrac{\varpi_{d}%
^{ps}\left(  D,\mathbf{X};\boldsymbol{\beta}\right)  \left(  Y\left(
d\right)  -\mathbb{E}[Y(d)]\right)  }{d\cdot p\left(  \mathbf{X}%
;\boldsymbol{\beta}\right)  +\left(  1-d\right)  \left(  1-p\left(
\mathbf{X};\boldsymbol{\beta}\right)  \right)  }\cdot\dot{p}\left(
\mathbf{X};\boldsymbol{\beta}\right)  \right\Vert \right]  <\infty,
\]
and $\left(  iii\right)  $ for some $\varepsilon>0$, $0<a_{1}<a_{2}<1$,
$F_{Y\left(  d\right)  }$ is continuously differentiable on $\left[
q_{Y\left(  d\right)  }\left(  a_{1}\right)  -\varepsilon,q_{Y\left(
d\right)  }\left(  a_{2}\right)  +\varepsilon\right]  $ with strictly positive
derivative $f_{Y\left(  d\right)  }$.
\end{assumption}

Assumption \ref{regularity}$(i)$ requires potential outcomes to be
square-integrable, whereas Assumption \ref{regularity}$(ii)$ is a mild
regularity condition which guarantees that, in a small neighborhood of
$\boldsymbol{\beta}_{0}$, the score of the IPW estimator for the ATE is
bounded by an integrable function. Assumption \ref{regularity}$(iii)$ requires
potential outcomes to be continuously distributed and only plays a role in the
analysis of quantile treatment effects. In principle, Assumption
\ref{regularity}$(iii)$ can be relaxed at the cost of using more complex
arguments, see \cite{Chernozhukov2017c} for details.

Before stating the results as a theorem, let us define some important
quantities. Let%
\begin{align}
\psi_{w,\Psi}^{ate}\left(  Y,D,\mathbf{X}\right)   &  =g^{ate}\left(
Y,D,\mathbf{X}\right)  -l_{w,\Psi}\left(  D,\mathbf{X};\boldsymbol{\beta}%
_{0}\right)  ^{\prime}\cdot\mathbf{G}_{\boldsymbol{\beta}}^{ate}%
,\label{infl.ate}\\
\psi_{w,\Psi}^{dte}\left(  Y,D,\mathbf{X};y\right)   &  =g^{dte}\left(
Y,D,\mathbf{X};y\right)  -l_{w,\Psi}\left(  D,\mathbf{X};\boldsymbol{\beta
}_{0}\right)  ^{\prime}\cdot\mathbf{G}_{\boldsymbol{\beta}}^{dte}\left(
y\right)  ,\label{infl.dte}\\
\psi_{w,\Psi}^{qte}\left(  Y,D,\mathbf{X};\tau\right)   &  =-\left(
g^{qte}\left(  Y,D,\mathbf{X};\tau\right)  -l_{w,\Psi}\left(  D,\mathbf{X}%
;\boldsymbol{\beta}_{0}\right)  ^{\prime}\cdot\mathbf{G}_{\boldsymbol{\beta}%
}^{qte}\left(  \tau\right)  \right)  \label{infl.qte}%
\end{align}
where, for $j\in\left\{  ate,dte,qte\right\}  $, $g^{j}\left(  Y,D,\mathbf{X}%
\right)  =g_{1}^{j}\left(  Y,D,\mathbf{X}\right)  -g_{0}^{j}\left(
Y,D,\mathbf{X}\right)  $, with%
\begin{align*}
g_{d}^{ate}\left(  Y,D,\mathbf{X}\right)   &  =\varpi_{d}^{ps}\left(
D,\mathbf{X};\boldsymbol{\beta}_{0}\right)  \cdot\left(  Y-\mathbb{E}\left[
Y\left(  d\right)  \right]  \right)  ,\\
g_{d}^{dte}\left(  Y,D,\mathbf{X};y\right)   &  =\varpi_{d}^{ps}\left(
D,\mathbf{X};\boldsymbol{\beta}_{0}\right)  \cdot\left(  1\left\{  Y\leq
y\right\}  -F_{Y\left(  d\right)  }\left(  y\right)  \right)  ,\\
g_{d}^{qte}\left(  Y,D,\mathbf{X};\tau\right)   &  =\frac{\varpi_{d}%
^{ps}\left(  D,\mathbf{X};\boldsymbol{\beta}_{0}\right)  \cdot\left(
1\left\{  Y\leq q_{Y\left(  d\right)  }\left(  \tau\right)  \right\}
-\tau\right)  }{f_{Y\left(  d\right)  }\left(  q_{Y\left(  d\right)  }\left(
\tau\right)  \right)  },
\end{align*}
and
\begin{align*}
\mathbf{G}_{\boldsymbol{\beta}}^{ate}  &  =\mathbb{E}\left[  \left(
\frac{g_{1}^{ate}\left(  Y,D,\mathbf{X}\right)  }{p\left(  \mathbf{X}%
;\boldsymbol{\beta}_{0}\right)  }+\frac{g_{0}^{ate}\left(  Y,D,\mathbf{X}%
\right)  }{1-p\left(  \mathbf{X};\boldsymbol{\beta}_{0}\right)  }\right)
\cdot\dot{p}\left(  \mathbf{X};\boldsymbol{\beta}_{0}\right)  \right]  ,\\
\mathbf{G}_{\boldsymbol{\beta}}^{dte}\left(  y\right)   &  =\mathbb{E}\left[
\left(  \frac{g_{1}^{dte}\left(  Y,D,\mathbf{X};y\right)  }{p\left(
\mathbf{X};\boldsymbol{\beta}_{0}\right)  }+\frac{g_{0}^{dte}\left(
Y,D,\mathbf{X};y\right)  }{1-p\left(  \mathbf{X};\boldsymbol{\beta}%
_{0}\right)  }\right)  \cdot\dot{p}\left(  \mathbf{X};\boldsymbol{\beta}%
_{0}\right)  \right]  ,\\
\mathbf{G}_{\boldsymbol{\beta}}^{qte}\left(  \tau\right)   &  =\mathbb{E}%
\left[  \left(  \frac{g_{1}^{qte}\left(  Y,D,\mathbf{X};\tau\right)
}{p\left(  \mathbf{X};\boldsymbol{\beta}_{0}\right)  }+\frac{g_{0}%
^{qte}\left(  Y,D,\mathbf{X};\tau\right)  }{1-p\left(  \mathbf{X}%
;\boldsymbol{\beta}_{0}\right)  }\right)  \cdot\dot{p}\left(  \mathbf{X}%
;\boldsymbol{\beta}_{0}\right)  \right]  .
\end{align*}
The functions $g^{ate}$, $g^{dte}$ and $g^{qte}$ would be the influence
functions of the ATE, DTE and QTE estimators, respectively, if the PS
parameters $\boldsymbol{\beta}_{0}$ were known. With some abuse of notation,
denote $\Omega_{w,\Psi}^{ate}=\mathbb{E}\left[  \psi_{w,\Psi}^{ate}\left(
Y,D,\mathbf{X}\right)  ^{2}\right]  $, $\Omega_{w,\Psi,y}^{dte}=\mathbb{E}%
\left[  \psi_{w,\Psi}^{dte}\left(  Y,D,\mathbf{X};y\right)  ^{2}\right]  $,
and $\Omega_{w,\Psi,\tau}^{qte}=\mathbb{E}\left[  \psi_{w,\Psi}^{qte}\left(
Y,D,\mathbf{X};\tau\right)  ^{2}\right]  $.

\begin{theorem}
\label{th.causal}Under Assumptions \ref{ignor} - \ref{regularity}, for each
$y\in\mathbb{R}$, $\tau\in\left[  \varepsilon,1-\varepsilon\right]  $, we have
that, as $n\rightarrow\infty$,%
\begin{align*}
\sqrt{n}\left(  \widehat{ATE}_{n}^{ips}-ATE\right)   &
\overset{d}{\rightarrow}N\left(  0,\Omega_{w,\Psi}^{ate}\right)  ,\\
\sqrt{n}\left(  \widehat{DTE}_{n}^{ips}-DTE\right)  \left(  y\right)   &
\overset{d}{\rightarrow}N\left(  0,\Omega_{w,\Psi,y}^{dte}\right)  ,\\
\sqrt{n}\left(  \widehat{QTE}_{n}^{ips}-QTE\right)  \left(  \tau\right)   &
\overset{d}{\rightarrow}N\left(  0,\Omega_{w,\Psi,\tau}^{qte}\right)  .
\end{align*}

\end{theorem}

Theorem \ref{th.causal} indicates that one can use our proposed IPS estimator
to estimate a variety of causal parameters that are able to highlight
treatment effect heterogeneity\footnote{Although the results stated in Theorem
\ref{th.causal} for distribution and quantile treatment effects are pointwise,
in Appendix \ref{main-results} we prove their uniform counterpart using
empirical process techniques. We omit the details in the main text only to
avoid additional cumbersome notation. We refer interested readers to the proof
of Theorem \ref{th.causal} in Appendix \ref{main-results} for additional
details.}. Furthermore, Theorem \ref{th.causal} also suggests that to conduct
asymptotically valid inference for these causal parameters, one simply needs
to estimate the asymptotic variance $\Omega_{w,\Psi}^{ate}$, $\Omega
_{w,\Psi,y}^{dte}$, and $\Omega_{w,\Psi,\tau}^{qte}$. Under additional
smoothness conditions (for instance, the PS being twice continuously
differentiable with bounded second derivatives), one can show that their
sample analogues are consistent using standard arguments. We omit the details
for the sake of brevity.\bigskip

\begin{Rem}
\label{att-rem}In Supplemental Appendix \ref{causal2}, we show that results
analogous to Theorem \ref{th.causal} also hold for the average, distributional
and quantile treatment effect on the treated. These treatment effects
parameters can have higher policy relevancy in setups where the policy
intervention is directed at individuals with certain characteristics, e.g.,
when a clinical treatment is directed to units with a specific symptoms; see
e.g., \cite{Heckman1997}.
\end{Rem}

\section{The IPS when treatment is endogenous\label{sec:endog}}

In many important applications, the assumption that treatment adoption is
exogenous may be too restrictive. For instance, when individuals do not comply
with their treatment assignment, or more generally when they sort into
treatment based on expected gains, Assumption \ref{ignor} is likely to be
violated. \cite{Imbens1994} and \cite{Angrist1996} point out that when this is
the case and a binary instrument ($Z)$ for the selection into treatment is
available, one can only nonparametrically identify treatment effect measures
for the subpopulation of compliers, that is, individuals who comply with their
actual assignment of treatment, and would have complied with the alternative
assignment. As shown by \cite{Abadie2003}, \cite{Frolich2007}, and
\cite{Frolich2013}, the instrument propensity score $q\left(  \mathbf{X}%
\right)  \equiv$ $\mathbb{P}(Z=1|X)$ plays a prominent role in this local
treatment effect (LTE) setup. In this section, we show that one can use the
IPS approach to estimate the instrument propensity score $q\left(
\mathbf{X}\right)  $, by maximizing covariate distribution balancing among
different instrument-by-treatment subgroups.

Before providing the details about how we apply the IPS approach to estimate
$q\left(  \mathbf{X}\right)  $ under the LTE setup, we introduce a brief
description of the LTE setup. Let $Z$ be a binary instrumental variable $Z$
for the treatment assignment. Denote $D\left(  0\right)  $ and $D\left(
1\right)  $ the value that $D$ would have taken if $Z$ is equal to zero or
one, respectively. The realized treatment is $D=ZD\left(  1\right)  +\left(
1-Z\right)  D\left(  0\right)  $. Thus, the observed sample in the LTE setup
consists of independent and identically distributed copies $\left\{  \left(
Y_{i},D_{i},Z_{i},\mathbf{X}_{i}^{^{\prime}}\right)  ^{\prime}\right\}
_{i=1}^{n}$. To identify the average, distributional and quantile treatment
effects for the compliers, we follow \cite{Abadie2003} and make the following assumption.

\begin{assumption}
\label{ass:late}($i$) $\left(  Y\left(  0\right)  ,Y\left(  1\right)
,D\left(  0\right)  ,D\left(  1\right)  \right)
\independent
Z|\mathbf{X}$; ($ii$) for some $\varepsilon>0$, $\varepsilon\leq q\left(
\mathbf{X}\right)  \leq1-\varepsilon$ $a.s.$ and $\mathbb{P}\left(  D\left(
1\right)  =1|\mathbf{X}\right)  >\mathbb{P}\left(  D\left(  0\right)
=1|\mathbf{X}\right)  $ $a.s.$; and ($iii$) $\mathbb{P}\left(  D\left(
1\right)  \geq D\left(  0\right)  |\mathbf{X}\right)  \allowbreak=1$ $a.s.$.
\end{assumption}

Assumption \ref{ass:late}($i$) imposes that once we condition on $\mathbf{X}$,
$Z$ is \textquotedblleft as good as randomly assigned\textquotedblright.
Assumption \ref{ass:late}($ii$) imposes a common support condition, and
guarantees that, conditional on $\mathbf{X}$, $Z$ is a relevant instrument for
$D$. Finally, Assumption \ref{ass:late}($iii$) is a monotonicity condition
that rules out the existence of defiers.

From \cite{Abadie2003} and \cite{Frolich2013}, we have that under Assumption
\ref{ass:late}, the average, distributional and quantile treatment effects for
compliers are nonparametrically identified, i.e.,%
\begin{align*}
LATE  &  \equiv\mathbb{E}\left[  Y\left(  1\right)  -Y\left(  0\right)
|\mathcal{C}\right]  =\mathbb{E}\left[  \varpi_{1}^{lte}\left(
D,Z,\mathbf{X;}q\right)  \cdot Y\right]  -\mathbb{E}\left[  \varpi_{0}%
^{lte}\left(  D,Z,\mathbf{X;}q\right)  \cdot Y\right]  ,\\
LDTE\left(  y\right)   &  \equiv\mathbb{P}\left(  Y\left(  1\right)  \leq
y|\mathcal{C}\right)  -\mathbb{P}\left(  Y\left(  0\right)  \leq
y|\mathcal{C}\right)  =F_{\varpi_{1}^{lte}\cdot Y}\left(  y\right)
-F_{\varpi_{0}^{lte}\cdot Y}\left(  y\right)  ,\\
LQTE\left(  \tau\right)   &  \equiv q_{Y\left(  1\right)  |\mathcal{C}}\left(
\tau\right)  -q_{Y\left(  0\right)  |\mathcal{C}}\left(  \tau\right)
=F_{\varpi_{1}^{lte}\cdot Y}^{-1}\left(  \tau\right)  -F_{\varpi_{0}%
^{lte}\cdot Y}^{-1}\left(  \tau\right)  ,
\end{align*}
where $\mathcal{C}$ denotes the complier subpopulation, and, for $d\in\left\{
0,1\right\}  $,
\begin{align}
\varpi_{d}^{lte}\left(  D,Z,\mathbf{X;}q\right)   &  =\frac{1\left\{
D=d\right\}  }{\kappa_{d}\left(  q\right)  }\left(  \frac{Z}{q\left(
\mathbf{X}\right)  }-\frac{\left(  1-Z\right)  }{1-q\left(  \mathbf{X}\right)
}\right)  ,\label{et}\\
F_{\varpi_{d}^{lte}\cdot Y}\left(  y\right)   &  =\mathbb{E}\left[  \varpi
_{d}^{lte}\left(  D,Z,\mathbf{X;}q\right)  \cdot1\left\{  Y\leq y\right\}
\right]  , \label{dt}%
\end{align}
and\vspace{-15pt}
\[
\kappa_{d}\left(  q\right)  \equiv\mathbb{E}\left[  \frac{1\left\{
D=d\right\}  Z}{q\left(  \mathbf{X}\right)  }-\frac{1\left\{  D=d\right\}
\left(  1-Z\right)  }{1\mathbb{-}q\left(  \mathbf{X}\right)  }\right]  ,
\]
and $F_{\varpi_{d}^{lte}\cdot Y}^{-1}\left(  \tau\right)  =\inf\left\{
y:F_{\varpi_{d}^{lte}\cdot Y}\left(  y\right)  \geq\tau\right\}  .$ From the
above results, it is clear that the instrument PS plays a prominent role in
the LTE setup, and that once we have an estimator for $q$ available, it is
relatively straightforward to construct estimators for the LATE, LDTE, and LQTE.

To estimate the instrument PS $q$, we adopt a parametric approach, i.e., we
assume that $q\left(  \mathbf{X}\right)  =q\left(  \mathbf{X;}%
\boldsymbol{\beta}_{0}^{lte}\right)  $, where $q$ is known up to the
finite-dimensional parameters $\boldsymbol{\beta}_{0}^{lte}$. Here, as we are
interested in treatment effects for the (latent) subpopulation of compliers,
we will attempt to estimate $\boldsymbol{\beta}_{0}^{lte}$ by maximizing the
covariate distribution balance among compliers. To do so, we build on Theorem
3.1 of \cite{Abadie2003}, which establishes that, for every measurable and
integrable function $f\left(  \mathbf{X}\right)  $ of the covariates
$\mathbf{X}$,%
\begin{align}
\mathbb{E}\left[  \varpi_{1}^{lte}\left(  D,Z,\mathbf{X;}\boldsymbol{\beta
}_{0}^{lte}\right)  \cdot f\left(  \mathbf{X}\right)  \right]   &
=\mathbb{E}\left[  \varpi^{lte}\left(  D,Z,\mathbf{X;}\boldsymbol{\beta}%
_{0}^{lte}\right)  \cdot f\left(  \mathbf{X}\right)  \right]  ,\nonumber\\
\mathbb{E}\left[  \varpi_{0}^{lte}\left(  D,Z,\mathbf{X;}\boldsymbol{\beta
}_{0}^{lte}\right)  \cdot f\left(  \mathbf{X}\right)  \right]   &
=\mathbb{E}\left[  \varpi^{lte}\left(  D,Z,\mathbf{X;}\boldsymbol{\beta}%
_{0}^{lte}\right)  \cdot f\left(  \mathbf{X}\right)  \right]  ,
\label{inst.cov.bal}%
\end{align}
where $\varpi_{d}^{lte}\left(  D,Z,\mathbf{X;}\boldsymbol{\beta}_{0}%
^{lte}\right)  $ is defined as in (\ref{et}) but with $\boldsymbol{\beta}%
_{0}^{lte}$ playing the role of $q$, as we assume that $q$ is a parametric
model, and%
\[
\varpi^{lte}\left(  D,Z,\mathbf{X;}\boldsymbol{\beta}\right)  =\frac{1}%
{\kappa\left(  \boldsymbol{\beta}\right)  }\left(  1-\frac{\left(  1-D\right)
Z}{q\left(  \mathbf{X;}\boldsymbol{\beta}\right)  }-\frac{D\left(  1-Z\right)
}{1-q\left(  \mathbf{X;}\boldsymbol{\beta}\right)  }\right)  ,
\]
with%
\[
\kappa\left(  \boldsymbol{\beta}\right)  \equiv\mathbb{E}\left[
1-\frac{\left(  1-D\right)  Z}{q\left(  \mathbf{X;}\boldsymbol{\beta}\right)
}-\frac{D\left(  1-Z\right)  }{1-q\left(  \mathbf{X;}\boldsymbol{\beta
}\right)  }\right]  .
\]
As noted in Theorem 3.1 of \cite{Abadie2003}, under Assumption \ref{ass:late},
$\mathbb{E}\left[  \varpi^{lte}\left(  D,Z,\mathbf{X;}\boldsymbol{\beta}%
_{0}^{lte}\right)  \cdot f\left(  \mathbf{X}\right)  \right]  =\mathbb{E}%
\left[  f\left(  \mathbf{X}\right)  |\mathcal{C}\right]  $, implying that
(\ref{inst.cov.bal}) are indeed balancing conditions for the complier subpopulation.

Next and analogously to the discussion in Section \ref{sec:ips}, we rewrite
(\ref{inst.cov.bal}) as%
\begin{equation}
\mathbf{H}_{w}^{lte}(\boldsymbol{\beta}_{0}^{lte},\mathbf{u})=\mathbf{0}%
~a.e~in~\mathbf{u}\in\Pi, \label{inst.cov.bal2}%
\end{equation}
where $\mathbf{H}_{w}^{lte}(\boldsymbol{\beta},\mathbf{u})=\mathbb{E}\left[
\mathbf{h}^{lte}\left(  D,Z,\mathbf{X};\boldsymbol{\beta}\right)
w(\mathbf{X};\mathbf{u})\right]  $, with $\mathbf{h}^{lte}\left(
D,Z,\mathbf{X};\boldsymbol{\beta}\right)  =(h_{1}^{lte}\left(  D,Z,\mathbf{X}%
;\boldsymbol{\beta}\right)  $,\allowbreak\ $h_{0}^{lte}\left(  D,Z,\mathbf{X}%
;\boldsymbol{\beta}\right)  )^{\prime}$, and, for $d\in\left\{  0,1\right\}
,$ $h_{d}^{lte}\left(  D,Z,\mathbf{X};\boldsymbol{\beta}\right)  =\varpi
_{d}^{lte}\left(  D,Z,\mathbf{X;}\boldsymbol{\beta}\right)  -\varpi
^{lte}\left(  D,Z,\mathbf{X;}\boldsymbol{\beta}\right)  $.

Based on (\ref{inst.cov.bal2}), we then show in Lemma \ref{lemma-C1} in the
Supplemental Appendix that $\boldsymbol{\beta}_{0}^{lte}$ is be globally
identified, i.e., $\boldsymbol{\beta}_{0}^{lte}$ is the unique minimizer of
the population minimum distance criteria $Q_{w}^{lte}(\boldsymbol{\beta}%
)=\int_{\Pi}\left\Vert \mathbf{H}_{w}^{lte}(\boldsymbol{\beta},\mathbf{u}%
)\right\Vert \,\Psi(d\mathbf{u})$. Thus, like in the case where treatment is
exogenous, we can fully exploit the balancing conditions (\ref{inst.cov.bal})
and estimate $\boldsymbol{\beta}_{0}^{lte}$ by
\begin{equation}
\widehat{\boldsymbol{\beta}}_{n,w}^{lips}=\arg\min_{\boldsymbol{\beta}%
\in\Theta}\int_{\Pi}\left\Vert \mathbf{H}_{n,w}^{lte}(\boldsymbol{\beta
},\mathbf{u})\right\Vert ^{2}\,\Psi_{n}(d\mathbf{u}), \label{lips}%
\end{equation}
where $\Psi_{n}$ is a uniformly consistent estimator of $\Psi$, $\mathbf{H}%
_{n,w}^{lte}(\boldsymbol{\beta},\mathbf{u})=\mathbb{E}\left[  \mathbf{h}%
_{n}^{lte}\left(  D,Z,\mathbf{X};\boldsymbol{\beta}\right)  w(\mathbf{X}%
;\mathbf{u})\right]  $, $\mathbf{h}_{n}^{lte}\left(  D,Z,\mathbf{X}%
;\boldsymbol{\beta}\right)  =(h_{n,1}^{lte}\left(  D,Z,\mathbf{X}%
;\boldsymbol{\beta}\right)  $, $h_{n,0}^{lte}\left(  D,Z,\mathbf{X}%
;\boldsymbol{\beta}\right)  )^{\prime}$, $h_{n,d}^{lte}\left(  D,Z,\mathbf{X}%
;\boldsymbol{\beta}\right)  =\varpi_{n,d}^{lte}\left(  D,Z,\mathbf{X;}%
\boldsymbol{\beta}\right)  -\varpi_{n}^{lte}\left(  D,Z,\mathbf{X;}%
\boldsymbol{\beta}\right)  $, and
\begin{align*}
\varpi_{n,d}^{lte}\left(  D,Z,\mathbf{X;}\boldsymbol{\beta}\right)   &
=\frac{1\left\{  D=d\right\}  }{\kappa_{n,d}\left(  \boldsymbol{\beta}\right)
}\left(  \frac{Z}{q\left(  \mathbf{X;}\boldsymbol{\beta}\right)  }%
-\frac{\left(  1-Z\right)  }{1-q\left(  \mathbf{X;}\boldsymbol{\beta}\right)
}\right) \\
\kappa_{n,d}\left(  \boldsymbol{\beta}\right)   &  =\mathbb{E}_{n}\left[
1\left\{  D=d\right\}  \left(  \frac{Z}{q\left(  \mathbf{X;}\boldsymbol{\beta
}\right)  }-\frac{1-Z}{1\mathbb{-}q\left(  \mathbf{X;}\boldsymbol{\beta
}\right)  }\right)  \right] \\
\varpi_{n}^{lte}\left(  D,Z,\mathbf{X;}\boldsymbol{\beta}\right)   &
=\frac{1}{\kappa_{n}\left(  \boldsymbol{\beta}\right)  }\left(  1-\frac
{\left(  1-D\right)  Z}{q\left(  \mathbf{X;}\boldsymbol{\beta}\right)  }%
-\frac{D\left(  1-Z\right)  }{1-q\left(  \mathbf{X;}\boldsymbol{\beta}\right)
}\right)  ,\\
\kappa_{n}\left(  \boldsymbol{\beta}\right)   &  =\mathbb{E}_{n}\left[
1-\frac{\left(  1-D\right)  Z}{q\left(  \mathbf{X;}\boldsymbol{\beta}\right)
}-\frac{D\left(  1-Z\right)  }{1-q\left(  \mathbf{X;}\boldsymbol{\beta
}\right)  }\right]  .
\end{align*}
As before, we focus our attention on the three weighting functions described
in Assumption \ref{ass.w}. We call (\ref{lips}) the local integrated
propensity score (LIPS) estimator.

In what follows, we derive the asymptotic properties of the instrument IPS
estimator $\widehat{\boldsymbol{\beta}}_{n,w}^{lips}$. Let the score of
$\mathbf{H}_{w}^{lte}(\boldsymbol{\beta},\mathbf{u})$ be defined as
$\mathbf{\dot{H}}_{w}^{lte}(\boldsymbol{\beta},\mathbf{u})=\left(
\mathbf{\dot{H}}_{1,w}^{lte^{\prime}}(\boldsymbol{\beta},\mathbf{u}%
),\mathbf{\dot{H}}_{0,w}^{lte^{\prime}}(\boldsymbol{\beta},\mathbf{u})\right)
^{\prime}$ where, for $d\in\left\{  0,1\right\}  ,$ $\mathbf{\dot{H}}%
_{d,w}^{lte}(\boldsymbol{\beta},\mathbf{u})=\mathbb{E}\left[  \mathbf{\dot{h}%
}_{d}^{lte}\left(  D,Z,\mathbf{X};\boldsymbol{\beta}\right)  w(\mathbf{X}%
;\mathbf{u})\right]  $,
\[
\mathbf{\dot{h}}_{d}^{lte}\left(  D,Z,\mathbf{X};\boldsymbol{\beta}\right)
=\boldsymbol{\dot{\varpi}}_{d}^{lte}\left(  D,Z,\mathbf{X;}\boldsymbol{\beta
}\right)  -\boldsymbol{\dot{\varpi}}^{lte}\left(  D,Z,\mathbf{X;}%
\boldsymbol{\beta}\right)  ,
\]
with \vspace{-15pt}
\begin{multline}
\boldsymbol{\dot{\varpi}}_{d}^{lte}\left(  D,Z,\mathbf{X;}\boldsymbol{\beta
}\right)  =-\frac{1\left\{  D=d\right\}  }{\kappa_{d}\left(  \boldsymbol{\beta
}\right)  }\left(  \frac{Z}{q\left(  \mathbf{X;}\boldsymbol{\beta}\right)
^{2}}+\frac{\left(  1-Z\right)  }{\left(  1-q\left(  \mathbf{X;}%
\boldsymbol{\beta}\right)  \right)  ^{2}}\right)  \cdot\dot{q}\left(
\mathbf{X;}\boldsymbol{\beta}\right)  ^{\prime}\nonumber\\
+\varpi_{d}^{lte}\left(  D,Z,\mathbf{X;}\boldsymbol{\beta}\right)
\cdot\mathbb{E}\left[  \frac{1\left\{  D=d\right\}  }{\kappa_{d}\left(
\boldsymbol{\beta}\right)  }\left(  \frac{Z}{q\left(  \mathbf{X;}%
\boldsymbol{\beta}\right)  ^{2}}+\frac{\left(  1-Z\right)  }{\left(
1-q\left(  \mathbf{X;}\boldsymbol{\beta}\right)  \right)  ^{2}}\right)
\cdot\dot{q}\left(  \mathbf{X;}\boldsymbol{\beta}\right)  ^{\prime}\right]  ,
\end{multline}
and \vspace{-15pt}
\begin{multline}
\boldsymbol{\dot{\varpi}}^{lte}\left(  D,Z,\mathbf{X;}\boldsymbol{\beta
}\right)  =-\frac{1}{\kappa\left(  \boldsymbol{\beta}\right)  }\left(
\frac{D\left(  1-Z\right)  }{\left(  1-q\left(  \mathbf{X;}\boldsymbol{\beta
}\right)  \right)  ^{2}}-\frac{\left(  1-D\right)  Z}{q\left(  \mathbf{X;}%
\boldsymbol{\beta}\right)  ^{2}}\right)  \cdot\dot{q}\left(  \mathbf{X;}%
\boldsymbol{\beta}\right)  ^{\prime}\nonumber\\
+\varpi^{lte}\left(  D,Z,\mathbf{X;}\boldsymbol{\beta}\right)  \cdot
\mathbb{E}\left[  \frac{1}{\kappa\left(  \boldsymbol{\beta}\right)  }\left(
\frac{D\left(  1-Z\right)  }{\left(  1-q\left(  \mathbf{X;}\boldsymbol{\beta
}\right)  \right)  ^{2}}-\frac{\left(  1-D\right)  Z}{q\left(  \mathbf{X;}%
\boldsymbol{\beta}\right)  ^{2}}\right)  \cdot\dot{q}\left(  \mathbf{X;}%
\boldsymbol{\beta}\right)  ^{\prime}\right]  ,
\end{multline}
and $\dot{q}\left(  \mathbf{\cdot;}\boldsymbol{\beta}\right)  =\left.  \left.
\partial q\left(  \mathbf{\cdot;}\boldsymbol{b}\right)  \right/
\partial\boldsymbol{b}\right\vert _{\boldsymbol{b=\beta}}$. We make the
following set of assumptions, which are the analogue of Assumption
\ref{ass.pscore}.

\begin{assumption}
\label{ass.int.pscore}$\left(  i\right)  q\left(  \mathbf{x}\right)  =q\left(
\mathbf{x};\boldsymbol{\beta}_{0}^{lte}\right)  $, where $\boldsymbol{\beta
}_{0}^{lte}$ is an interior point of a compact set $\Theta\subset
\mathbb{R}^{k};$ $\left(  ii\right)  $ for some $\delta>0$, $\delta\leq
q\left(  \mathbf{x;}\boldsymbol{\beta}\right)  \leq1-\delta$ for all
$\mathbf{x}\in\mathcal{X}$, $\boldsymbol{\beta}\in int\left(  \Theta\right)
$; $\left(  iii\right)  $ with probability one, $q\left(  \mathbf{X;}%
\boldsymbol{\beta}\right)  $ is continuous at each $\boldsymbol{\beta}\in$
$\Theta$; $\left(  iv\right)  $ with probability one, $q\left(  \mathbf{X;}%
\boldsymbol{\beta}\right)  $ is continuously differentiable in a neighborhood
of $\boldsymbol{\beta}_{0}^{lte}$, $\Theta_{0}^{lte}$ $\subset\Theta$ $;$
$\left(  v\right)  $ for $d\in\left\{  0,1\right\}  $
\[
\mathbb{E}\left[  \sup_{\boldsymbol{\beta}\in\Theta_{0}^{lte}}\left\Vert
1\left\{  D=d\right\}  \left(  \frac{Z}{q\left(  \mathbf{X;}\boldsymbol{\beta
}\right)  ^{2}}+\frac{\left(  1-Z\right)  }{\left(  1-q\left(  \mathbf{X;}%
\boldsymbol{\beta}\right)  \right)  ^{2}}\right)  \cdot\dot{q}\left(
\mathbf{X;}\boldsymbol{\beta}\right)  \right\Vert \right]  <\infty.
\]

\end{assumption}

The next theorem characterizes the asymptotic properties of the instrument IPS
estimators $\widehat{\boldsymbol{\beta}}_{n,w}^{lips}$. Define the $k\times k$
matrix\vspace{-5pt}
\[
C_{w,\Psi}^{lte}=\int_{\Pi}\left(  \mathbf{\dot{H}}_{w}^{lte}%
(\boldsymbol{\beta}_{0}^{lte},\mathbf{u})^{c}~\mathbf{\dot{H}}_{w}%
^{lte}(\boldsymbol{\beta}_{0}^{lte},\mathbf{u})+\mathbf{\dot{H}}_{w}%
^{lte}(\boldsymbol{\beta}_{0}^{lte},\mathbf{u})^{\prime}\left(  \mathbf{\dot
{H}}_{w}^{lte}(\boldsymbol{\beta}_{0}^{lte},\mathbf{u})^{\prime}\right)
^{c}\right)  \Psi(d\mathbf{u}),
\]
and the $k\times1$ vector \vspace{-15pt}
\begin{multline}
l_{w,\Psi}^{lte}\left(  D,Z,\mathbf{X};\boldsymbol{\beta}_{0}^{lte}\right)
=-\left(  C_{w,\Psi}^{lte}\right)  ^{-1}\cdot\int_{\Pi}\left(  \mathbf{\dot
{H}}_{w}^{lte}(\boldsymbol{\beta}_{0}^{lte},\mathbf{u})^{c}~w(\mathbf{X}%
;\mathbf{u})+\mathbf{\dot{H}}_{w}^{lte}(\boldsymbol{\beta}_{0}^{lte}%
,\mathbf{u})^{\prime}w(\mathbf{X};\mathbf{u})^{c}\right)  \Psi(d\mathbf{u})\\
\cdot\mathbf{h}^{lte}\left(  D,Z,\mathbf{X};\boldsymbol{\beta}_{0}%
^{lte}\right)  . \label{lin.rep.inst}%
\end{multline}

\begin{theorem}
\label{th.inst.ips}Under Assumptions \ref{ass.w}, \ref{ass:late}, and
\ref{ass.int.pscore}, as $n\rightarrow\infty$,%
\[
\widehat{\boldsymbol{\beta}}_{n,w}^{lips}-\boldsymbol{\beta}_{0}^{lte}%
=o_{p}\left(  1\right)  .
\]
Furthermore, provided that the matrix $C_{w,\Psi}^{lte}$ is positive definite,%
\[
\sqrt{n}\left(  \widehat{\boldsymbol{\beta}}_{n,w}^{lips}-\boldsymbol{\beta
}_{0}^{lte}\right)  =\frac{1}{\sqrt{n}}\sum_{i=1}^{n}l_{w,\Psi}^{lte}\left(
D_{i},Z_{i},\mathbf{X}_{i};\boldsymbol{\beta}_{0}^{lte}\right)  +o_{p}\left(
1\right)  ,
\]
and%
\[
\sqrt{n}\left(  \widehat{\boldsymbol{\beta}}_{n,w}^{lips}-\boldsymbol{\beta
}_{0}^{lte}\right)  \overset{d}{\rightarrow}N\left(  0,\Omega_{w,\Psi}%
^{lips}\right)  \text{,}%
\]
where $\Omega_{w,\Psi}^{lips}\equiv\mathbb{E}\left[  l_{w,\Psi}^{lte}\left(
D,\mathbf{X};\boldsymbol{\beta}_{0}^{lte}\right)  l_{w,\Psi}^{lte}\left(
D,\mathbf{X};\boldsymbol{\beta}_{0}^{lte}\right)  ^{\prime}\right]  .$
\end{theorem}

With the results of Theorem \ref{th.inst.ips} at hand, we can estimate the
LATE, LDTE, and LQTE by using the instrument IPS estimators:%
\begin{align}
\widehat{LATE}_{n}^{lips}  &  =\mathbb{E}_{n}\left[  \left(  \varpi
_{n,1}^{lte}\left(  D,Z,\mathbf{X};\widehat{\boldsymbol{\beta}}_{n,w}%
^{lips}\right)  -\varpi_{n,0}^{lte}\left(  D,Z,\mathbf{X}%
;\widehat{\boldsymbol{\beta}}_{n,w}^{lips}\right)  \right)  Y\right]
,\label{late.ips}\\
\widehat{LDTE}_{n}^{lips}\left(  y\right)   &  =\widehat{F^{r}}_{n,\varpi
_{1}^{lte}\cdot Y}\left(  y\right)  -\widehat{F^{r}}_{n,\varpi_{0}^{lte}\cdot
Y}\left(  y\right)  ,\label{ldte.ips}\\
\widehat{LQTE}_{n}^{lips}\left(  \tau\right)   &  =\widehat{F^{r}}%
_{n,\varpi_{1}^{lte}\cdot Y}^{-1}\left(  \tau\right)  -\widehat{F^{r}%
}_{n,\varpi_{0}^{lte}\cdot Y}^{-1}\left(  \tau\right)  , \label{lqte.ips}%
\end{align}
where, for $d\in\left\{  0,1\right\}  $, $\widehat{F^{r}}_{n,\varpi_{d}%
^{lte}\cdot Y}$ $\left(  \cdot\right)  $ denotes the rearrangement of
$\widehat{F}_{n,~\varpi_{d}^{lte}\cdot Y}\left(  \cdot\right)  $,%
\[
\widehat{F}_{n,\varpi_{d}^{lte}\cdot Y}\left(  \cdot\right)  =\mathbb{E}%
_{n}\left[  \varpi_{n,d}^{lte}\left(  D,Z,\mathbf{X}%
;\widehat{\boldsymbol{\beta}}_{n,w}^{lips}\right)  1\left\{  Y\leq
\cdot\right\}  \right]  \text{, }%
\]
if $\widehat{F}_{n,\varpi_{d}^{lte}\cdot Y}$ is not monotone, see, e.g.,
\cite{Chernozhukov2010}, and \cite{Wuthrich2019}\footnote{Lack of monotonicity
may appear in finite samples because the weights $w_{n,d}^{lte}$ can be
negative. This poses problems for the inversion of the weighted cumulative
distribution functions to obtain the quantile functions. On the other hand,
under Assumption \ref{ass:late}, the population weights $w_{d}^{lte}$ are
non-negative, implying that these potential problems disappear,
asymptotically. As discussed in detail in \cite{Chernozhukov2010}, we can
bypass such challenges by monotonizing $\widehat{F}_{n,~w_{d}^{lte}\cdot Y}$
via rearrangements.}. Importantly, these rearrangements do not change the
asymptotic properties of the estimators.

To derive the asymptotic properties of (\ref{late.ips})-(\ref{lqte.ips}), we
impose the following regularity conditions, which are the analogue of
Assumption \ref{regularity}.

\begin{assumption}
\label{ass.inst.regularity}For $d\in\left\{  0,1\right\}  $, $\left(
i\right)  \mathbb{E}\left[  Y\left(  d\right)  ^{2}|\mathcal{C}\right]  <M$
for some $0<M<\infty$, $\left(  ii\right)  $
\[
\mathbb{E}\left[  \sup_{\beta\in\Theta_{0}^{lte}}\left\Vert 1\left\{
D=d\right\}  \left(  Y\left(  d\right)  -\mathbb{E}[\left.  Y(d)\right\vert
\mathcal{C}]\right)  \left(  \frac{Z}{q\left(  \mathbf{X;}\boldsymbol{\beta
}\right)  ^{2}}+\frac{\left(  1-Z\right)  }{\left(  1-q\left(  \mathbf{X;}%
\boldsymbol{\beta}\right)  \right)  ^{2}}\right)  \cdot\dot{q}\left(
\mathbf{X;}\boldsymbol{\beta}\right)  \right\Vert \right]  <\infty,
\]
and $\left(  iii\right)  $ for some $\varepsilon>0$, $0<a_{1}<a_{2}<1$,
$F_{Y\left(  d\right)  |\mathcal{C}}$ is continuously differentiable on
$\left[  q_{Y\left(  d\right)  |\mathcal{C}}\left(  a_{1}\right)
-\varepsilon,q_{Y\left(  d\right)  |\mathcal{C}}\left(  a_{2}\right)
+\varepsilon\right]  $ with strictly positive derivative $f_{Y\left(
d\right)  |\mathcal{C}}$.
\end{assumption}

\begin{theorem}
\label{th.inst.causal}Under Assumptions \ref{ass.w}, \ref{ass:late}%
-\ref{ass.inst.regularity}, for each $y\in\mathbb{R}$, $\tau\in\left[
\varepsilon,1-\varepsilon\right]  $, we have that, as $n\rightarrow\infty$,%
\begin{align*}
\sqrt{n}\left(  \widehat{LATE}_{n}^{lips}-LATE\right)   &
\overset{d}{\rightarrow}N\left(  0,\Omega_{w,\Psi}^{late}\right)  ,\\
\sqrt{n}\left(  \widehat{LDTE}_{n}^{lips}-LDTE\right)  \left(  y\right)   &
\overset{d}{\rightarrow}N\left(  0,\Omega_{w,\Psi,y}^{ldte}\right)  ,\\
\sqrt{n}\left(  \widehat{LQTE}_{n}^{lips}-LQTE\right)  \left(  \tau\right)
&  \overset{d}{\rightarrow}N\left(  0,\Omega_{w,\Psi,\tau}^{lqte}\right)  ,
\end{align*}
where $\Omega_{w,\Psi}^{late}$, $\Omega_{w,\Psi,y}^{ldte}$ and $\Omega
_{w,\Psi,\tau}^{lqte}$ are defined in the proof of Theorem
\ref{th.inst.causal} in Appendix \ref{main-results-inst}.\bigskip
\end{theorem}

\begin{Rem}
Although the results stated in Theorem \ref{th.inst.causal} for local
distribution and quantile treatment effects are pointwise, in Appendix
\ref{main-results-inst} we prove their uniform counterpart using empirical
process techniques. We omit the details in the main text only to avoid
additional cumbersome notation. We refer interested readers to the proof of
Theorem \ref{th.inst.causal} in Appendix \ref{main-results-inst} for
additional details.
\end{Rem}

\begin{Rem}
For brevity, we focused on the unconditional LATE, LDTE and LQTE causal
parameters. However, we would like to mention that one can readily use the
instrument IPS discussed in this section to estimate other conditional
treatment effect measures, such as the conditional local quantile treatment
effects introduced by \cite{Abadie2002a}, and the local average response
functions introduced by \cite{Abadie2003}. Given the results in Theorem
\ref{th.inst.ips}, establishing the asymptotic properties of these conditional
treatment effect measures is relatively straightforward.
\end{Rem}

\begin{Rem}
We note that under Assumption \ref{ass:late}, when one fixes $f(\mathbf{X}%
)=\mathbf{X}$ and subtracts the second equality in (\ref{inst.cov.bal}) from
the first equality in (\ref{inst.cov.bal}), one has that, after some
straightforward manipulation,
\[
\mathbb{E}\left[  \left(  \frac{Z}{q\left(  \mathbf{X};\boldsymbol{\beta}%
_{0}^{lte}\right)  }-\frac{1-Z}{1-q\left(  \mathbf{X};\boldsymbol{\beta}%
_{0}^{lte}\right)  }\right)  \mathbf{X} \right]  =\mathbf{0}
.\label{cbps.inst}%
\]
Thus, by substituting $D$ and $p\left(  \mathbf{X};\boldsymbol{\beta}%
_{0}\right)  $ in (\ref{cbps}) with $Z$ and $q\left(  \mathbf{X}%
;\boldsymbol{\beta}_{0}^{lte}\right)  $, one can, in principle, use
\cite{Imai2014}'s covariate balancing propensity score procedure to estimate
the instrument propensity score. However (and analogous to the discussion in
Section \ref{sec2}), such a procedure would only partly exploit Theorem 3.1 of
\cite{Abadie2003}, which is in contrast with our proposed LIPS procedure. As a
consequence, the LIPS estimation procedure can lead to estimators with
improved finite-sample properties; we illustrate this point via Monte Carlo
simulations in Section \ref{sec:MC2}.
\end{Rem}

\section{Monte Carlo simulations\label{simu}}

\subsection{Unconfoundedness setup\label{simu.exog}\medskip}

In this section, we conduct a series of Monte Carlo experiments to study the
finite sample properties of our proposed treatment effect estimators based on
the IPS. We first compare the performance of different IPW estimators for the
ATE and the QTE$\left(  \tau\right)  $, $\tau\in\left\{
0.25,0.5,0.75\right\}  $ when one estimates the PS using our proposed IPS
estimators (\ref{ips-ind})-(\ref{ips-exp}), the classical maximum likelihood
(ML) approach, \cite{Imai2014}'s just-identified covariate balancing
propensity score (CBPS) as in (\ref{cbps}) with $f\left(  \mathbf{X}\right)
=\mathbf{X}$, and \cite{Imai2014}'s overidentified CBPS (\ref{cbps}) with
$f\left(  \mathbf{X}\right)  =\left(  \mathbf{X}^{\prime},\dot{p}\left(
\mathbf{X};\boldsymbol{\beta}\right)  ^{\prime}\right)  ^{\prime}$, i.e., on
top of balancing the means, one also makes use of the likelihood score
equation. In all cases, we consider a logistic PS model where all available
covariates enter linearly. All treatment effect estimators use stabilized
weights (\ref{w1}) and (\ref{w0}).

We consider sample size $n$ equal to $500$\footnote{Simulation results with
$n=200$ and $n=1000$ lead to similar conclusions and are available on
request.}. For each design, we conduct $1,000$ Monte Carlo simulations. We
compare the various IPW estimators in terms of average bias, root mean square
error (RMSE), relative mean square error (relMSE), empirical 95\% coverage
probability, the median length of a 95\% confidence interval, and the
asymptotic relative efficiency (ARE)\footnote{For any parameter $\eta$ of a
distribution $F$, and for estimators $\widehat{\eta}_{1}$ and $\widehat{\eta
}_{2}$ approximately $N\left(  \eta,V_{1}/n\right)  $ and $N\left(  \eta
,V_{2}/n\right)  $, respectively, the asymptotic relative efficiency of
$\widehat{\eta}_{2}$ with respect to $\widehat{\eta}_{1}$ is given by
$V_{1}/V_{2}$; see, e.g., Section 8.2 in \cite{VanderVaart1998}. Thus, to
compute the ARE for our estimators, we build on Theorem \ref{th.causal} and
replace the asymptotic variances with their sample analogues.}. For the
relative measures of performance, relMSE and ARE, we treat estimators based on
the overidentified CBPS as the benchmark. The confidence intervals are based
on the normal approximation in Theorem \ref{th.causal}, with the asymptotic
variances being estimated by their sample analogues. For the variance of
QTE$\left(  \tau\right)  $ estimators, we estimate the potential outcome
densities using the Gaussian kernel coupled with Silverman's rule-of-thumb
bandwidth - these are the default choices of the \texttt{density} function in
the \texttt{stats} package in \texttt{R}. We use the \texttt{CBPS} package in
\texttt{R }to estimate both CBPS estimators. Finally, we emphasize that our
measures of performance highlight not only the behavior of IPW point estimates
but also the accuracy of their associated inference procedures.

Our simulation design is largely based on \cite{Kang2007}. Let $\mathbf{X}%
=\left(  X_{1},X_{2},\right.  $ $\left.  X_{3},X_{4}\right)  ^{\prime}$ be
distributed as $N\left(  0,I_{4}\right)  $, and $I_{4}$ be the $4\times4$
identity matrix. The true PS is given by%
\[
p\left(  \mathbf{X}\right)  =\frac{\exp\left(  -X_{1}+0.5X_{2}-0.25X_{3}%
-0.1X_{4}\right)  }{1+\exp\left(  -X_{1}+0.5X_{2}-0.25X_{3}-0.1X_{4}\right)
},
\]
and the treatment status $D$ is generated as $D=1\left\{  p\left(
\mathbf{X}\right)  >U\right\}  $, where $U$ follows a uniform $\left(
0,1\right)  $ distribution. The potential outcomes $Y\left(  1\right)  $ and
$Y\left(  0\right)  $ are given by%
\begin{align*}
Y\left(  1\right)   &  =210+m\left(  \mathbf{X}\right)  +\varepsilon\left(
1\right)  , & Y\left(  0\right)   &  =200-m\left(  \mathbf{X}\right)
+\varepsilon\left(  0\right)  ,
\end{align*}
where $m\left(  \mathbf{X}\right)  =27.4X_{1}+13.7X_{2}+13.7X_{3}+13.7X_{4}$,
$\varepsilon\left(  1\right)  $ and $\varepsilon\left(  0\right)  $ are
independent $N\left(  0,1\right)  $ random variables. The ATE and the
QTE$\left(  \tau\right)  $ are equal to 10, for all $\tau\in\left(
0,1\right)  $.

We consider two different scenarios to assess the sensibility of the proposed
estimators under misspecified models that are \textquotedblleft nearly
correct\textquotedblright. In the first experiment, the observed data is
$\left\{  \left(  Y_{i},D_{i},\mathbf{X}_{i}^{^{\prime}}\right)  ^{\prime
}\right\}  _{i=1}^{n}$, and, therefore, all IPW estimators are correctly
specified. In the second experiment the observed data is $\left\{  \left(
Y_{i},D_{i},\mathbf{W}_{i}^{^{\prime}}\right)  ^{\prime}\right\}  _{i=1}^{n}$,
where $\mathbf{W}=\left(  W_{1},W_{2},W_{3},W_{4}\right)  ^{\prime}$ with
$W_{1}=\exp\left(  X_{1}/2\right)  $, $W_{2}=X_{2}/\left(  1+\exp\left(
X_{1}\right)  \right)  ,$ $W_{3}=\left(  X_{1}X_{3}/25+0.6\right)  ^{3}$, and
$W_{4}=\left(  X_{2}+X_{4}+20\right)  ^{2}$. In this second scenario, the IPW
estimators for ATE and QTE$\left(  \tau\right)  $ are misspecified.

Table \ref{tab:sim.tab} displays the simulation results for both scenarios.
When the PS model is correctly specified, all estimators perform well in terms
of bias and coverage probability, i.e., all estimators are essentially
unbiased and their associated confidence intervals have correct coverage.
Comparing ML-based with CBPS-based estimators, we note that IPW estimators
based on ML tend to have higher mean square error, longer confidence
intervals, and lower ARE. Thus, it is clear that CBPS-based IPW estimators can
improve upon those based on ML. However, our simulation results under correct
specification suggest that we can improve further the performance of the CBPS
estimator by fully exploiting the covariate balancing of the propensity score.
For instance, the relative mean square error of estimators based on the IPS
with either projection or exponential weight function tend to be at least 10\%
smaller than those based on the CBPS, with the exception of the QTE$\left(
0.25\right)  $. The gains in terms of ARE also tend to be large. For example,
the ARE of the ATE estimator based on the IPS with projection weight function
with respect to the one based on the overidentified CBPS is 1.26. This implies
that the ATE estimator based on the overidentified CBPS would require
$1.26\times n$ observations to perform equivalently to the ATE estimator based
on IPS with projection weight. IPS estimators based on the exponential weight
also tend to dominate CBPS estimators in terms of mean square errors and ARE.
Finally, we note that IPW estimators based on the IPS with the indicator
function tend to give slightly larger confidence intervals than when using
other IPS estimators, perhaps because there are multiple covariates (four in
our simulation design), implying that many $1\left\{  \mathbf{X}_{i}%
\leq\mathbf{u}\right\}  $ are equal to zero when $\mathbf{u}$ is evaluated at
the sample observations.

\begin{table}[H]
\caption{Monte Carlo study of the performance of IPW estimators for ATE and
$QTE$ based on different propensity score estimation methods. Sample size:
$n=500$.}%
\label{tab:sim.tab}%
\centering
%
\par
\begin{adjustbox}{ max width=1\linewidth, max totalheight=1\textheight, keepaspectratio}
\begin{threeparttable}
\begin{tabular}{@{}lrrrrrrrrrrrrr@{}} \toprule
& \multicolumn{6}{c}{Correctly Specified Model} &  \hphantom{a}  & \multicolumn{6}{c}{Misspecified Model} \\
	\cmidrule{2-7}  \cmidrule{8-14}
& \multicolumn{1}{c}{Bias} &  \multicolumn{1}{c}{RMSE} & \multicolumn{1}{c}{relMSE} & \multicolumn{1}{c}{COV}  & \multicolumn{1}{c}{ACIL} & \multicolumn{1}{c}{ARE} &
\hphantom{a} &
\multicolumn{1}{c}{Bias} &  \multicolumn{1}{c}{RMSE} & \multicolumn{1}{c}{relMSE} & \multicolumn{1}{c}{COV}  & \multicolumn{1}{c}{ACIL} & \multicolumn{1}{c}{ARE} \\
	
\noalign{\vskip 3mm} $(a)$ ${\bf{ATE}}$ \\
\noalign{\vskip 1mm}\hphantom{ab} $IPS_{exp}$ &    	  0.091 & 3.669 & 0.885 & 0.944 & 14.068 &   1.216 &   & 1.889 & 4.157 & 0.729 & 0.909 & 13.792 &  1.378 \\
\noalign{\vskip 1mm}\hphantom{ab} $IPS_{ind}$& 	 0.966 & 3.659 & 0.880 & 0.966 & 15.556 &   0.995 &   & 2.533 & 4.743 & 0.949 & 0.955 & 17.798 &    0.827 \\
\noalign{\vskip 1mm}\hphantom{ab} $IPS_{proj}$& 	  0.091 & 3.603 & 0.853 & 0.942 & 13.830 &    1.259 &   & 0.387 & 3.527 & 0.525 & 0.965 & 15.105 &    1.149 \\
\noalign{\vskip 1mm}\hphantom{ab} $CBPS_{just}$  &		      0.080 & 4.023 & 1.064 & 0.941 & 14.983   & 1.072 &   & 2.736 & 4.729 & 0.943 & 0.857 & 13.922   & 1.352 \\
\noalign{\vskip 1mm}\hphantom{ab} $CBPS_{over}$  &       0.071 & 3.900 & 1.000 & 0.960 & 15.515   & 1.000 &   & 2.673 & 4.869 & 1.000 & 0.918 & 16.190 &    1.000 \\
\noalign{\vskip 1mm}\hphantom{ab} $MLE$   &  		    0.092 & 4.371 & 1.256 & 0.945 & 16.221 &  0.915 &   & 6.444 & 12.280 & 6.361 & 0.836 & 20.755 &    0.608 \\
\noalign{\vskip 3mm} $(b)$ $\bf{QTE(0.25)}$ \\
\noalign{\vskip 1mm}\hphantom{ab} $IPS_{exp}$ &      -0.015 & 4.380 & 1.061 & 0.954 & 17.373 &   1.035 &   & -2.211 & 4.936 & 1.202 & 0.917 & 17.205 &   1.067 \\
\noalign{\vskip 1mm}\hphantom{ab} $IPS_{ind}$ &      0.557 & 4.625 & 1.183 & 0.971 & 19.473 &    0.824 &   & -1.140 & 4.760 & 1.118 & 0.959 & 19.816 &   0.804 \\
\noalign{\vskip 1mm}\hphantom{ab} $IPS_{proj}$ &  -0.001 & 4.372 & 1.057 & 0.951 & 17.340 &   1.039 &   & -1.490 & 4.759 & 1.117 & 0.983 & 23.364 &    0.578 \\
\noalign{\vskip 1mm}\hphantom{ab} $CBPS_{just}$ &  			    -0.022 & 4.350 & 1.047 & 0.956 & 17.209   & 1.054 &   & -1.311 & 4.580 & 1.035 & 0.938 & 17.160   & 1.072 \\
\noalign{\vskip 1mm}\hphantom{ab} $CBPS_{over}$  &  			   -0.062 & 4.252 & 1.000 & 0.966 & 17.672   & 1.000 &   & -1.128 & 4.502 & 1.000 & 0.948 & 17.769    & 1.000 \\
\noalign{\vskip 1mm}\hphantom{ab} $MLE$  &			    -0.055 & 4.403 & 1.072 & 0.960 & 17.567&   1.012 &   & 1.376 & 10.837 & 5.793 & 0.948 & 20.934 &    0.720 \\
\noalign{\vskip 3mm} $(c)$ $\bf{QTE(0.50)}$ \\
\noalign{\vskip 1mm}\hphantom{ab} $IPS_{exp}$&        0.032 & 4.266 & 0.936 & 0.957 & 17.724 &   1.135 &   & 0.986 & 4.472 & 0.834 & 0.955 & 17.439 &    1.210 \\
\noalign{\vskip 1mm}\hphantom{ab} $IPS_{ind}$ &       0.829 & 4.408 & 0.999 & 0.972 & 19.301 &   0.957 &   & 1.762 & 4.895 & 1.000 & 0.958 & 20.217 &   0.900 \\
\noalign{\vskip 1mm}\hphantom{ab} $IPS_{proj}$&       0.010 & 4.234 & 0.922 & 0.955 & 17.562 &   1.156 &   & 0.030 & 4.279 & 0.764 & 0.971 & 18.573 &   1.067 \\
\noalign{\vskip 1mm}\hphantom{ab} $CBPS_{just}$ & 			 0.068 & 4.582 & 1.080 & 0.956 & 18.543 &   1.037 &   & 1.914 & 4.887 & 0.996 & 0.928 & 17.802   & 1.161 \\
\noalign{\vskip 1mm}\hphantom{ab} $CBPS_{over}$  & 			 0.003 & 4.409 & 1.000 & 0.972 & 18.879    & 1.000 &   & 1.834 & 4.896 & 1.000 & 0.951 & 19.185   & 1.000 \\
\noalign{\vskip 1mm}\hphantom{ab} $MLE$ &      0.076 & 4.758 & 1.165 & 0.963 & 19.396 &   0.947 &   & 5.936 & 14.363 & 8.606 & 0.912 & 25.292 &   0.575 \\
\noalign{\vskip 3mm} $(d)$ $\bf{QTE(0.75)}$ \\
\noalign{\vskip 1mm}\hphantom{ab} $IPS_{exp}$ &      -0.001 & 5.701 & 0.940 & 0.935 & 21.887 &   1.149 &   & 5.340 & 7.588 & 0.826 & 0.828 & 21.017    & 1.350 \\
\noalign{\vskip 1mm}\hphantom{ab} $IPS_{ind}$ &    1.222 & 5.431 & 0.853 & 0.960 & 22.343 &   1.103 &   & 5.788 & 8.151 & 0.953 & 0.893 & 25.225 & 0.937 \\
\noalign{\vskip 1mm}\hphantom{ab} $IPS_{proj}$ &     0.021 & 5.611 & 0.911 & 0.938 & 21.474 &   1.194 &   & 2.100 & 5.442 & 0.425 & 0.968 & 24.135   & 1.024 \\
\noalign{\vskip 1mm}\hphantom{ab} $CBPS_{just}$ &      -0.012 & 6.229 & 1.122 & 0.935 & 23.455    & 1.001 &   & 6.374 & 8.648 & 1.073 & 0.777 & 21.506   & 1.289 \\
\noalign{\vskip 1mm}\hphantom{ab} $CBPS_{over}$ &      -0.012 & 5.880 & 1.000 & 0.952 & 23.461   & 1.000 &   & 5.955 & 8.351 & 1.000 & 0.861 & 24.418   & 1.000 \\
\noalign{\vskip 1mm}\hphantom{ab} $MLE$ &   -0.004 & 6.627 & 1.270 & 0.938 & 25.097   & 0.874 &   & 11.915 & 19.011 & 5.182 & 0.754 & 31.666 &  0.595 \\
\bottomrule
\end{tabular}%
\begin{tablenotes}[para,flushleft]
\footnotesize{
Note: Simulations based on 1,000 Monte Carlo experiments. Bias, Monte Carlo Bias; RMSE, Monte Carlo root mean square error; relMSE, relative Monte Carlo mean square error; COV, Monte Carlo coverage of 95\% normal confidence interval; ACIL, Monte Carlo average of 95\% normal confidence interval length;
ARE, asymptotic relative efficiency; ATE, average treatment effect; QTE($\tau$), quantile treatment effect at $\tau$ quantile. Both relMSE and ARE are expressed with respect to the IPW estimator based on the overidentified CBPS.  The propensity score model is based on a logistic link function. $IPS_{ind}$, IPW estimator based on IPS estimator (\ref{ips-ind});
$IPS_{proj}$, IPW estimator based on IPS estimator (\ref{ips-proj});
$IPS_{exp}$, IPW estimator based on IPS estimator (\ref{ips-exp});
$CBPS_{just}$, IPW estimator based on the (just-identified) CBPS estimator with moment equation (\ref{cbps}), with $f\left(\mathbf{X}\right) = \mathbf{X}$;
$CBPS_{over}$ , IPW estimator based on the (overidentified) CBPS estimator with moment equation (\ref{cbps}), with $f\left(\mathbf{X}\right)  =\left(  \mathbf{X}^{\prime}, \dot{p}\left(
\mathbf{X};\boldsymbol{\beta}\right)  ^{\prime}\right)  ^{\prime}$, with $\dot{p}\left(
\mathbf{X};\boldsymbol{\beta}\right) $ the derivative of the propensity score model with respect to $\boldsymbol{\beta}$;
$MLE$ , IPW estimator based on MLE. }
\end{tablenotes}
\end{threeparttable}
\end{adjustbox}\end{table}

When the PS model is misspecified, our Monte Carlo results suggest that the
potential gains of using the IPS can also be pronounced. In this scenario, we
note that estimators based on ML tend to be substantially biased, have
relatively high RMSE, and inference tends to be misleading. These findings are
in line with the results in \cite{Kang2007}. Overall, estimators based on
just-identified CBPS improve on ML, though under-coverage is still an
unresolved issue when one focuses on the ATE and QTE$\left(  0.75\right)  $.
Estimators based on the overidentified CBPS tend to have better coverage than
those based on the just-identified CBPS, but under-coverage of QTE$\left(
0.75\right)  $ is still severe, perhaps because of the large biases. Finally,
we note that our proposed IPS estimators tend to further improve upon CBPS. In
particular, estimators based on the IPS with the projection weight function
have the lowest bias and RMSE, and their confidence intervals are close to the
nominal coverage --- the only exception is when one focuses on QTE$\left(
0.25\right)  $, where estimators based on CBPS tends to perform slightly
better than our proposed IPS procedure. On the other hand, we note that, in
terms of mean square error, the gains of adopting the IPS estimator with
either projection or exponential weighting function tend to be large in all
other considered causal measures, especially for ATE and QTE$\left(
0.75\right)  $.

\subsection{Local Treatment Effect Setup\label{sec:MC2}}

We now consider the setup where treatment is endogenous but one has access to
a binary instrument $Z$, as described in Section \ref{sec:endog}. Here, we
compare the performance of different IPW estimators for the LATE and the
LQTE$\left(  \tau\right)  $, $\tau\in\left\{  0.25,0.5,0.75\right\}  $ when
one estimates the instrument PS $q\left(  \cdot\right)  $ using our proposed
instrument IPS estimator (\ref{lips}) with exponential, indicator and
projection-based weights, the classical ML approach, \cite{Imai2014}'s
just-identified and overidentified CBPS with $Z$ playing the role of $D$. In
all cases, we consider a logistic instrument PS model where all available
covariates enter linearly. As in the unconfoundedness case, we consider sample
size $n$ equal to $500$, and conduct $1,000$ Monte Carlo simulations for each design.

The simulation design is similar to the one in Section \ref{simu.exog}. Let
$\mathbf{X}$, $\mathbf{W,}$ $Y\left(  1\right)  $, and $Y\left(  0\right)  $
be defined as before. The true instrument PS is given by
\[
q\left(  \mathbf{X}\right)  =\frac{\exp\left(  -X_{1}+0.5X_{2}-0.25X_{3}%
-0.1X_{4}\right)  }{1+\exp\left(  -X_{1}+0.5X_{2}-0.25X_{3}-0.1X_{4}\right)
},
\]
the instrument $Z$ is generated as $Z=1\left\{  q\left(  \mathbf{X}\right)
>U_{1}\right\}  $, where $U_{1}$ follows a uniform $\left(  0,1\right)  $
distribution. The potential treatments $D\left(  1\right)  $ and $D\left(
0\right)  $ are generated as $D\left(  1\right)  =1\{p^{\ast}\left(  Y\left(
1\right)  -Y\left(  0\right)  \right)  >U_{2}\}$ and $D\left(  0\right)  =0,$
where $U_{2}$ follows a uniform $\left(  0,1\right)  $ distribution, and%
\[
p^{\ast}\left(  Y\left(  1\right)  -Y\left(  0\right)  \right)  =\frac
{\exp\left(  2+0.05\cdot\left(  Y\left(  1\right)  -Y\left(  0\right)
\right)  \right)  }{1+\exp\left(  2+0.05\cdot\left(  Y\left(  1\right)
-Y\left(  0\right)  \right)  \right)  }.
\]
Finally, the realized treatment is $D=Z\cdot D\left(  1\right)  +\left(
1-Z\right)  \cdot D\left(  0\right)  $, and the realized outcome is $Y=D\cdot
Y\left(  1\right)  +\left(  1-D\right)  \cdot Y\left(  0\right)  $. The LATE,
LQTE$\left(  0.25\right)  $, LQTE$\left(  0.5\right)  $, and LQTE$\left(
0.75\right)  $ are approximately equal to $39.25,$ 42.94, 35, and 42.94,
respectively. This design is consistent with a generalized Roy model, under
which individuals with higher treatment effects are more likely to be treated
if they are eligible for treatment. We also emphasize that, given the
one-sided non-compliance, LATE and LQTE are equal to the ATT and QTT in this scenario.

As before, we consider two scenarios. On the first one, the observed data is
$\left\{  \left(  Y_{i},D_{i},Z_{i},\mathbf{X}_{i}^{^{\prime}}\right)
^{\prime}\right\}  _{i=1}^{n}$, and, therefore, all IPW estimators are
correctly specified. In the second scenario, the observed data is $\left\{
\left(  Y_{i},D_{i},Z_{i},\mathbf{W}_{i}^{^{\prime}}\right)  ^{\prime
}\right\}  _{i=1}^{n}$, and all considered IPW estimators for LATE and
LQTE$\left(  \tau\right)  $ are misspecified.

Table \ref{tab:sim.endog.tab} displays the simulation results for both
scenarios. When the instrument PS model is correctly specified, all estimators
perform well in terms of bias and coverage probability, except the estimators
based on the LIPS estimator (\ref{lips}) with the indicator weighting function
--- the bias of the local treatment effect estimators based on LIPS with
indicator function is non-negligible when $n=500$, and such biases distort the
confidence intervals. In additional simulations, we note that the bias
associated with estimators based on the LIPS with the indicator weighting
function converges to zero when sample size grows, though the rate of
convergence is rather slow. As such, we recommend that, in practice, one
should favor the other PS estimators with respect to the LIPS with the
indicator weighting function. Like in the unconfoundedness setup, we note that
IPW estimators based on ML tend to have higher mean square error, longer
confidence intervals, and lower ARE than the IPW estimators based on the
just-identified CBPS estimator; the performance of the overidentified CBPS is,
in general, worse than MLE, specially for LATE. The results in Table
\ref{tab:sim.endog.tab} also show that, when the instrument propensity score
is correctly specified, the LIPS estimators with the exponential or projection
weighting function tend to outperform the other methods, particularly when
estimating the LATE and LQTE$\left(  0.75\right)  $.

When the instrument PS model is misspecified, our Monte Carlo results suggest
that using the LIPS can also be attractive. In this setup, we note that
estimators based on ML tend to have higher biases, RMSE and misleading
confidence intervals. Local treatment effect estimators based on the
(instrumented) CBPS improve upon those based on ML, with the
just-identified\ CBPS estimator performing better than the overidentified
CBPS. However, under-coverage is still an issue, except when one focuses on
LQTE$\left(  0.25\right)  $. On the other and, our simulation results suggest
that our proposed LIPS estimators lead to local treatment effect estimators
with even better statistical properties than those based on the (instrumented)
CBPS --- such gains are specially pronounced when estimating the local
treatment effect parameters based on the LIPS with the exponential or
projection weighting functions.

\begin{table}[H]
\caption{Monte Carlo study of the performance of IPW estimators for LATE and
$LQTE$ based on different instrument propensity score estimation methods.
Sample size: $n=500$.}%
\label{tab:sim.endog.tab}%
\centering
%
\par
\begin{adjustbox}{ max width=1\linewidth, max totalheight=1\textheight, keepaspectratio}
\begin{threeparttable}
\begin{tabular}{@{}llrrrrrrrrrrrr@{}} \toprule
& \multicolumn{6}{c}{Correctly Specified Model} &  \hphantom{a}  & \multicolumn{6}{c}{Misspecified Model} \\
	\cmidrule{2-7}  \cmidrule{8-14}
& \multicolumn{1}{c}{Bias} &  \multicolumn{1}{c}{RMSE} & \multicolumn{1}{c}{relMSE} & \multicolumn{1}{c}{COV}  & \multicolumn{1}{c}{ACIL} & \multicolumn{1}{c}{ARE} &
\hphantom{a} &
\multicolumn{1}{c}{Bias} &  \multicolumn{1}{c}{RMSE} & \multicolumn{1}{c}{relMSE} & \multicolumn{1}{c}{COV}  & \multicolumn{1}{c}{ACIL} & \multicolumn{1}{c}{ARE} \\
	
\noalign{\vskip 3mm} $(a)$ ${\bf{LATE}}$ \\
\noalign{\vskip 1mm} \hphantom{ab} $LIPS_{exp}$&    -0.253 & 4.420 & 0.782 & 0.956 & 17.784 &   1.746 &   & 5.132 & 6.645 & 0.356 & 0.938 & 21.586 &   1.299 \\
\noalign{\vskip 1mm}\hphantom{ab} $LIPS_{ind}$ & 	 -5.510 & 6.700 & 1.798 & 0.751 & 16.317   & 2.074 &   & -0.692 & 4.863 & 0.191 & 0.953 & 20.184 &   1.485 \\
\noalign{\vskip 1mm}\hphantom{ab} $LIPS_{proj}$ & 	 -1.010 & 4.325 & 0.749 & 0.955 & 17.058   & 1.897 &   & 0.392 & 4.764 & 0.183 & 0.987 & 27.773 &   0.784 \\
\noalign{\vskip 1mm}\hphantom{ab} $CBPS_{just}$ &		 	-0.051 & 4.723 & 0.893 & 0.939 & 17.710   & 1.760 &   & 8.038 & 9.683 & 0.756 & 0.592 & 18.421 &   1.783 \\
\noalign{\vskip 1mm}\hphantom{ab} $CBPS_{over}$ &    	 1.384 & 4.997 & 1.000 & 0.984 & 23.496 &   1.000 &   & 9.359 & 11.135 & 1.000 & 0.713 & 24.598 &   1.000 \\
\noalign{\vskip 1mm}\hphantom{ab} $MLE$  &  		 0.165 & 5.385 & 1.161 & 0.950 & 20.183 &  1.355 &  & 11.195 & 15.515 & 1.941 & 0.612 & 24.415 &   1.015 \\
\noalign{\vskip 3mm} $(b)$ $\bf{LQTE(0.25)}$ \\
\noalign{\vskip 1mm} \hphantom{ab} $LIPS_{exp}$&   -0.235 & 4.294 & 1.126 & 0.956 & 17.408 &   1.111 &   & -0.475 & 4.163 & 0.773 & 0.967 & 17.697 &    1.052 \\
\noalign{\vskip 1mm} \hphantom{ab} $LIPS_{ind}$&    -3.036 & 5.456 & 1.818 & 0.906 & 19.076 &    0.925 &   & -3.281 & 5.498 & 1.349 & 0.894 & 18.734 &    0.939 \\
\noalign{\vskip 1mm} \hphantom{ab} $LIPS_{proj}$&  -0.768 & 4.313 & 1.136 & 0.948 & 17.234 &    1.133 &   & -0.854 & 4.680 & 0.977 & 0.968 & 20.512 &    0.783 \\
\noalign{\vskip 1mm}\hphantom{ab} $CBPS_{just}$&  			 -0.074 & 4.051 & 1.002 & 0.959 & 16.845    & 1.186 &   & 1.288 & 4.414 & 0.869 & 0.951 & 16.794 &    1.168 \\
\noalign{\vskip 1mm}\hphantom{ab} $CBPS_{over}$ &  			 0.544 & 4.046 & 1.000 & 0.977 & 18.348    & 1.000 &   & 1.906 & 4.734 & 1.000 & 0.952 & 18.154 &   1.000 \\
\noalign{\vskip 1mm}\hphantom{ab} $MLE$ &			   0.044 & 4.167 & 1.060 & 0.961 & 17.376 &    1.115 &   & 3.685 & 11.064 & 5.461 & 0.932 & 20.920 &   0.753 \\
\noalign{\vskip 3mm} $(c)$ $\bf{LQTE(0.50)}$ \\
\noalign{\vskip 1mm} \hphantom{ab} $LIPS_{exp}$ &   -0.409 & 4.523 & 1.012 & 0.963 & 18.928 &   1.224    & & 1.782 & 4.911 & 0.509 & 0.966 & 19.785 &   1.145 \\
\noalign{\vskip 1mm} \hphantom{ab} $LIPS_{ind}$ &      -4.995 & 6.583 & 2.143 & 0.840 & 19.025 &   1.212   & & -2.334 & 5.335 & 0.600 & 0.935 & 19.906 &    1.131 \\
\noalign{\vskip 1mm} \hphantom{ab} $LIPS_{proj}$ &    -1.154 & 4.526 & 1.013 & 0.958 & 18.465 &   1.286   & & -0.634 & 4.842 & 0.495 & 0.976 & 22.156 &    0.913 \\
\noalign{\vskip 1mm}\hphantom{ab} $CBPS_{just}$ & 			-0.209 & 4.531 & 1.015 & 0.958 & 18.894   & 1.229 &   & 4.041 & 6.427 & 0.871 & 0.862 & 18.731 &  1.277 \\
\noalign{\vskip 1mm}\hphantom{ab} $CBPS_{over}$& 			 0.438 & 4.497 & 1.000 & 0.977 & 20.943   & 1.000 &   & 4.538 & 6.885 & 1.000 & 0.890 & 21.167 &   1.000 \\
\noalign{\vskip 1mm}\hphantom{ab} $MLE$&      -0.039 & 4.798 & 1.138 & 0.960 & 20.005 &  1.096 &   & 8.165 & 16.100 & 5.468 & 0.852 & 25.739   & 0.676 \\
\noalign{\vskip 3mm} $(d)$ $\bf{LQTE(0.75)}$ \\
\noalign{\vskip 1mm} \hphantom{ab} $LIPS_{exp}$ &    -0.381 & 5.741 & 0.941 & 0.973 & 24.263 &   1.328 &   & 5.186 & 7.680 & 0.477 & 0.922 & 25.923 &    1.240 \\
\noalign{\vskip 1mm} \hphantom{ab} $LIPS_{ind}$&   -7.576 & 9.143 & 2.386 & 0.729 & 21.698 &   1.661 &   & -1.153 & 6.274 & 0.319 & 0.949 & 25.504 &    1.281 \\
\noalign{\vskip 1mm} \hphantom{ab} $LIPS_{proj}$&   -1.230 & 5.613 & 0.899 & 0.964 & 23.285 &    1.442 &   & -0.048 & 5.890 & 0.281 & 0.984 & 33.207 &   0.756 \\
\noalign{\vskip 1mm}\hphantom{ab} $CBPS_{just}$&     -0.048 & 6.136 & 1.075 & 0.958 & 25.116    & 1.240 &   & 7.853 & 10.323 & 0.863 & 0.766 & 24.619   & 1.375 \\
\noalign{\vskip 1mm}\hphantom{ab} $CBPS_{over}$ &     0.874 & 5.919 & 1.000 & 0.981 & 27.964   & 1.000 &   & 8.568 & 11.114 & 1.000 & 0.819 & 28.867   & 1.000 \\
\noalign{\vskip 1mm}\hphantom{ab} $MLE$&  0.128 & 6.744 & 1.298 & 0.966 & 27.475 &   1.036 &  & 13.486 & 20.394 & 3.367 & 0.749 & 33.944 &   0.723 \\
\bottomrule
\end{tabular}%
\begin{tablenotes}[para,flushleft]
\footnotesize{
Note: Simulations based on 1,000 Monte Carlo experiments. Bias, Monte Carlo Bias; RMSE, Monte Carlo root mean square error; relMSE, relative Monte Carlo mean square error; COV, Monte Carlo coverage of 95\% normal confidence interval; ACIL, Monte Carlo average of 95\% normal confidence interval length;
ARE, asymptotic relative efficiency; LATE, local average treatment effect; LQTE($\tau$), local quantile treatment effect at $\tau$ quantile. Both relMSE and ARE are expressed with respect to the IPW estimator based on the overidentified CBPS.  All instrument propensity scores is based on a logistic link function. $LIPS_{ind}$, $LIPS_{proj}$ and $LIPS_{exp}$ are the IPW estimators based on LIPS estimator (\ref{lips}) with the indicator, projection, and exponential weight function, respectively;
$CBPS_{just}$, IPW estimator based on the (just-identified) CBPS estimator with moment equation (\ref{cbps}), with $Z$ in the place of $D$ and $f\left(\mathbf{X}\right) = \mathbf{X}$;
$CBPS_{over}$, IPW estimator based on the (overidentified) CBPS estimator with moment equation (\ref{cbps}),  with $Z$ in the place of $D$ and $f\left(\mathbf{X}\right)  =\left(  \mathbf{X}^{\prime}, \dot{p}\left(
\mathbf{X};\boldsymbol{\beta}\right)  ^{\prime}\right)  ^{\prime}$, with $\dot{p}\left(
\mathbf{X};\boldsymbol{\beta}\right) $ the derivative of the instrument propensity score model with respect to $\boldsymbol{\beta}$;
$MLE$, IPW estimator based on MLE. }
\end{tablenotes}
\end{threeparttable}
\end{adjustbox}\end{table}

Overall, our Monte Carlo simulations illustrate that, by fully exploiting the
covariate balancing property of the (instrument) PS, we can get treatment
effect estimators with improved finite sample properties. Our simulation
results also point out that treatment effect estimators based on the IPS and
LIPS estimators with either exponential or projection weighting functions tend
to perform better than when one uses the indicator weighting function. As
such, we recommend that, in practice, one should favor these weighting
functions with respect to the indicator weighting function, especially when
the dimension of the covariates included in the PS model is moderate or
high\footnote{In unreported additional simulations, we also have found that
the numerical performance of $IPS_{ind}$ and $LIPS_{ind}$ is sometimes
sensitive to initial values used in the optimization procedure when the number
of included covariates is moderate. We argue that this is additional reason to
favor the other weighting functions with respect to the indicator one.}.

\section{Empirical illustrations\label{sec:application}}

In this section, we apply our proposed tools to two different datasets. First,
we revisit \cite{Ichino2008} and use Italian data from the early 2000s to
study if temporary work agency (TWA) assignment affects the probability of
finding a stable job later on. Second, we study the effect of 401(k)
retirement plan on asset accumulation using data from the Survey of Income and
Program Participation, as in \cite{Benjamin2003a}, \cite{Abadie2003}, and
\cite{Chernozhukov2004}.

\subsection{Effect of temporary work assignment on future stable employment}

In temporary agency work, a company that needs employees signs a contract with
a TWA, which, in turn, is in charge of hiring and subsequently leasing these
workers to the company. In contrast to \textquotedblleft
traditional\textquotedblright\ jobs, the TWA is in charge of paying the
workers salary and fringe benefits, whereas the company's responsibility is to
train and guide the workers. One of the main arguments of introducing
temporary agency work is that it helps workers facing barriers to employment
find a stable job later on.

To evaluate whether TWA assignment has a positive impact on employment,
\cite{Ichino2008} collected data for two Italian regions, Tuscany and Sicily,
in the early 2000s. The dataset contains 2030 individuals, 511 of them treated
and 1519 untreated. Here, the treated group consists of individuals who were
on a TWA assignment during the first 6 months of 2001, whereas the untreated
group contains individuals aged 18 - 40, who belonged to the labor force but
did not have a stable job on January 2001, and who did not have a TWA
assignment during the first semester of 2001. Thus, both treatment groups were
drawn from the same local labor market. The outcome of interest is having a
permanent job at the end of 2002. A rich set of variables related to
demographic characteristics, family background, educational achievements, and
work experience before the treatment period were collected to adjust for
potential confounding (see Table 1 in \cite{Ichino2008}). Using PS matching,
\cite{Ichino2008} find evidence that TWA assignment has a positive effect on
permanent employment, especially in Tuscany. The results for Sicily are
sensitive to small violations of the strong ignorability assumption.
Therefore, in what follows, we focus on the Tuscany sub-sample\footnote{The
data are publicly available at
http://qed.econ.queensu.ca/jae/2008-v23.3/ichino-mealli-nannicini/.}.

\begin{table}[pth]
\caption{Treatment Effect of TWA assignment on the probability to find a
permanent job: IPW estimators for the ATE using different propensity score
estimation methods.}%
\label{tab:tempw}%
\centering
\par
\begin{adjustbox}{ max width=1\linewidth, max totalheight=1\textheight, keepaspectratio}
\begin{threeparttable}
\begin{tabular}{@{}llccccc}\toprule
\noalign{\vskip 2mm} & & \phantom{abc}$MLE$ & \phantom{ac}$CBPS_{just}$ & \phantom{ac}$CBPS_{over}$  & \phantom{abc}$IPS_{exp}$& \phantom{abc}$IPS_{proj}$ \\ \midrule
Whole Sample & & \multicolumn{1}{c}{\phantom{abc}17.83} & \multicolumn{1}{c}{\phantom{abc}20.67} & \multicolumn{1}{c}{\phantom{abc}17.95} &
\multicolumn{1}{c}{\phantom{abc}18.31} &
\multicolumn{1}{c}{\phantom{abc}18.03} \\
& &\multicolumn{1}{c}{\phantom{abc}(4.62)} & \multicolumn{1}{c}{\phantom{abc}(3.90)} & \multicolumn{1}{c}{\phantom{abc}(4.40)} &
\multicolumn{1}{c}{\phantom{abc}(3.53)} &
\multicolumn{1}{c}{\phantom{abc}(4.07)} \\
Male & & \multicolumn{1}{c}{\phantom{abc}14.40} & \multicolumn{1}{c}{\phantom{abc}22.79} & \multicolumn{1}{c}{\phantom{abc}18.33} &
\multicolumn{1}{c}{\phantom{abc}18.51} &
\multicolumn{1}{c}{\phantom{abc}18.64}\\
& & \multicolumn{1}{c}{\phantom{abc}(7.22)} & \multicolumn{1}{c}{\phantom{abc}(5.43)} &  \multicolumn{1}{c}{\phantom{abc}(5.89)} &
\multicolumn{1}{c}{\phantom{abc}(5.01)} &
\multicolumn{1}{c}{\phantom{abc}(5.38)}\\
Female & & \multicolumn{1}{c}{\phantom{abc}16.01} & \multicolumn{1}{c}{\phantom{abc}18.58} &\multicolumn{1}{c}{\phantom{abc}15.64} &
\multicolumn{1}{c}{\phantom{abc}15.40} &
\multicolumn{1}{c}{\phantom{abc}17.91}\\
& & \multicolumn{1}{c}{\phantom{abc}(5.64)} & \multicolumn{1}{c}{\phantom{abc}(5.95)} & \multicolumn{1}{c}{\phantom{abc}(6.30)} &
\multicolumn{1}{c}{\phantom{abc}(4.33)} &
\multicolumn{1}{c}{\phantom{abc}(4.45)} \\
\bottomrule
\end{tabular}%
\begin{tablenotes}[para,flushleft]
\footnotesize{
Note:  Same data used by \cite{Ichino2008}. The propensity score model is based on a logistic link function. Standard errors are in parentheses.
The estimators are the same as those we describe in Table \ref{tab:sim.tab}.}
\end{tablenotes}
\end{threeparttable}
\end{adjustbox}\end{table}

We use the results in Sections \ref{causal} to estimate the ATE. We compare
different IPW estimators based on the same PS estimation methods as in the
simulation studies in Section \ref{simu}, except the IPS coupled with the
indicator weighting function as it tends to be numerically unstable when dimension of covariates is moderate.
Table \ref{tab:tempw} shows the point estimates and standard errors (in
parentheses) for the whole Tuscany sample, and presents some heterogeneity
results based on gender. The PS specification we use is the one adopted by
\cite{Ichino2008}, which includes all the pre-treatment variables mentioned in
Table 1 of \cite{Ichino2008}, squared distance, and an interaction between
self-employment and one of the provinces.

The results in Table \ref{tab:tempw} suggest that the ATE is positive, and
statistically significant at the conventional levels, regardless of the
estimation procedure adopted. The overall average effect of TWA assignment on
the probability of having a permanent job ranges from 18 to 21, 14 to 23, and
15 to 19 percentage points when using the whole sample, the male
subpopulation, and the female subpopulation, respectively. Interestingly, the
IPS estimators can provide gains of efficiency when compared to both the MLE
and CBPS estimators. For instance, for the subsample of females, the
asymptotic relative efficiency (ARE) of the ATE estimator based on the IPS
with exponential, and projection weights with respect to the one based on MLE
are 1.70, and 1.58, respectively, while the ARE for the ATE based on the just
and overidentified CBPS with respect to the one based on MLE are,
respectively, 0.9 and 0.8. These findings suggest that the IPS can indeed lead
to improved treatment effect estimators in relevant settings.

\subsection{Effect of 401(k) retirement plans on asset accumulation}

As discussed in \cite{Benjamin2003a}, \cite{Abadie2003},
\cite{Chernozhukov2004}, and many others, tax-deferred retirement plans have
been popular in the US since the 1980s. A main goal of these programs is to
increase individual saving for retirement. Amongst the most popular
tax-deferred programs is the 401(k) plan. Interestingly, 401(k) plans are
provided by employers, and, therefore, only workers in firms that offer such
programs are eligible. On the other hand, we emphasize that \textit{eligible}
employees choose whether to participate (i.e., make a contribution) or not,
making the evaluation of the effectiveness of 401(k) plans on accumulated
assets more challenging as a result of endogeneity concerns --- individuals
who participate in 401(k) programs have stronger preferences for savings and
would have saved more even in the absence of these programs.

To bypass the endogeneity challenge, \cite{Benjamin2003a} uses data from the
1991 Survey of Income and Program Participation (SIPP) and compares households
that are eligible with those who are non-eligible for 401(k) plans to assess
the effect of \textit{eligibility} on accumulated assets. He argues that since
401(k) eligibility is determined by the employers, household preference for
savings plays a negligible role in determining eligibility once one controls
for observed household characteristics. Using PS matching,
\cite{Benjamin2003a} finds evidence that 401(k) eligibility has a positive
effect on asset accumulation.

\cite{Abadie2003}, \cite{Chernozhukov2004} and \cite{Wuthrich2019}, on the
other hand, study the effect of 401(k) \textit{participation} on asset
accumulation, using 401(k) eligibility as an instrument for the actual
participation status. Similarly to \cite{Benjamin2003a}, they argue that
401(k) eligibility is exogenous after controlling for a vector of observed
household characteristics. \cite{Abadie2003}, using a semiparametric IPW
estimator for the LATE, finds that the effect of 401(k) participation on net
financial assets is significant and positive. \cite{Chernozhukov2004} and
\cite{Wuthrich2019}, using an IV quantile regression model, also find positive
and significant effects of 401(k) participation on net financial assets.

In what follows, we apply the methodology discussed in Sections \ref{causal}
and \ref{sec:endog} to study the effects of eligibility and participation in
401(k) programs on saving behavior. As suggested by \cite{Benjamin2003a},
\cite{Abadie2003}, and \cite{Chernozhukov2004}, eligibility is assumed to be
exogenous after controlling for covariates. Also note that, because only
eligible individuals can enroll in 401(k) plans, the monotonicity condition in
Assumption \ref{ass:late}$\left(  iii\right)  $ holds trivially, and the LATE
and LQTE estimators presented in Section \ref{sec:endog} approximate the
average and quantile treatment effect for the treated (i.e., for 401(k) participants).

\begin{table}[ptbh]
\caption{Effects of 401(k) plan on different measures of wealth}%
\label{tab:asset1}
\centering
\par
\begin{adjustbox}{ max width=1\linewidth, max totalheight=1\textheight, keepaspectratio}
\begin{threeparttable}
\begin{tabular}{lccccccccccc}  \noalign{\vskip -3mm} \toprule
& \multicolumn{11}{c}{Panel A: Effects of 401(k) plan eligibility on wealth} \\   \noalign{\vskip 2mm}   \cmidrule{1-12}
& \multicolumn{5}{c}{Outcome: Net Financial Assets} &   \phantom{abc} & \multicolumn{5}{c}{Outcome: Total Wealth} \\  \cmidrule{2-6} \cmidrule{8-12}
\noalign{\vskip -1mm}& \multicolumn{1}{c}{$MLE$} & \multicolumn{1}{c}{$CBPS_{just}$} & \multicolumn{1}{c}{$CBPS_{over}$} & \multicolumn{1}{c}{$IPS_{exp}$} & \multicolumn{1}{c}{$IPS_{proj}$} &   &  \multicolumn{1}{c}{$MLE$} & \multicolumn{1}{c}{$CBPS_{just}$} & \multicolumn{1}{c}{$CBPS_{over}$} & \multicolumn{1}{c}{$IPS_{exp}$} & \multicolumn{1}{c}{$IPS_{proj}$} \\ \cmidrule{2-6} \cmidrule{8-12}
\bf{ATE} &           8,138  &           8,190  &           8,820  &           8,218  &           7,788  &   &           6,049  &           5,997  &           7,906  &           6,589  &           5,402  \\
			\noalign{\vskip -1mm}& (1,135) & (1,150) & (1,362) & (1,376) & (1,604) &   & (1,823) & (1,811) & (2,486) & (2,201) & (2,797) \\
\noalign{\vskip 1mm} \bf{QTE(0.25)}&              996  &              996  &           1,000  &           1,000  &              996  &   &           3,024  &           2,917  &           3,425  &           2,993  &           2,950  \\
			\noalign{\vskip -1mm}& (229) & (228) & (237) & (225) & (231) &   & (611) & (591) & (789) & (593) & (617) \\
\noalign{\vskip 1mm} \bf{QTE(0.50)} &           4,447  &           4,200  &           4,559  &           4,350  &           4,300  &   &           7,402  &           7,419  &           9,027  &           7,615  &           7,419  \\
			\noalign{\vskip -1mm} & (278) & (259) & (331) & (276) & (309) &   & (1,162) & (1,111) & (1,580) & (1,143) & (1,157) \\
\noalign{\vskip 1mm} \bf{QTE(0.75)} &         13,065  &         12,995  &         13,980  &         13,339  &         12,859  &   &           9,131  &           8,871  &         13,050  &         10,419  &           8,665  \\
			\noalign{\vskip -1mm}& (931) & (922) & (1,166) & (964) & (1,025) &   & (2,833) & (2,786) & (3,742) & (2,972) & (3,158) \\ \cmidrule{1-12}
\noalign{\vskip 2mm} & \multicolumn{11}{c}{Panel B: Effects of 401(k) plan participation on wealth} \\   \noalign{\vskip 2mm}  \cmidrule{1-12}
\noalign{\vskip 2mm}  & \multicolumn{5}{c}{Outcome: Net Financial Assets} &   & \multicolumn{5}{c}{Outcome: Total Wealth} \\  \cmidrule{2-6} \cmidrule{8-12}
\noalign{\vskip -1mm}&  \multicolumn{1}{c}{$MLE$} & \multicolumn{1}{c}{$CBPS_{just}$} & \multicolumn{1}{c}{$CBPS_{over}$} & \multicolumn{1}{c}{$LIPS_{exp}$} & \multicolumn{1}{c}{$LIPS_{proj}$}  &   &  \multicolumn{1}{c}{$MLE$} & \multicolumn{1}{c}{$CBPS_{just}$} & \multicolumn{1}{c}{$CBPS_{over}$} & \multicolumn{1}{c}{$LIPS_{exp}$} & \multicolumn{1}{c}{$LIPS_{proj}$} \\ \cmidrule{2-6} \cmidrule{8-12}
\bf{LATE} &         11,674  &         11,700  &         12,767  &         12,107  &         11,176  &   &           8,706  &           8,568  &         11,590  &           9,922  &           7,740  \\
			\noalign{\vskip -1mm} & (1,621) & (1,640) & (1,929) & (1,929) & (2,250) &   & (2,609) & (2,587) & (3,532) & (3,093) & (3,872) \\
\noalign{\vskip 1mm} \bf{LQTE(0.25)}	 &           1,618  &           1,536  &           1,753  &           1,589  &           1,529  &   &           5,226  &           4,853  &           6,204  &           5,200  &           5,003  \\
			\noalign{\vskip -1mm}	& (284) & (284) & (302) & (278) & (285) &   & (948) & (907) & (1,207) & (893) & (924) \\
\noalign{\vskip 1mm} \bf{LQTE(0.50)}	&           7,285  &           7,041  &           7,849  &           7,341  &           7,197  &   &        10,187  &           9,925  &         12,701  &         10,730  &        10,026  \\
	\noalign{\vskip -1mm} & (525) & (507) & (644) & (512) & (518) &   & (1,279) & (1,232) & (1,696) & (1,249) & (1,316) \\
\noalign{\vskip 1mm} \bf{LQTE(0.75)}	 &         19,939  &         19,589  &         21,772  &         20,325  &         19,410  &   &        14,061  &        13,200  &         19,909  &         16,353  &        13,041  \\
\noalign{\vskip -1mm} & (1,034) & (1,015) & (1,331) & (1,068) & (1,136) &   & (1,054) & (1,037) & (1,342) & (1,087) & (1,159) \\
\bottomrule
\end{tabular}%
\begin{tablenotes}[para,flushleft]
\footnotesize{
Note:  Same data used by \cite{Benjamin2003a} and \cite{Chernozhukov2004}. The propensity score model is based on a logistic link function. Standard errors in parentheses. The estimators in Panel A are
the same as those we describe in Table \ref{tab:sim.tab}, whereas those in Panel B are the same as those described in Table \ref{tab:sim.endog.tab}.}
\end{tablenotes}
\label{tab:addlabel}%
\end{threeparttable}
\end{adjustbox}
\end{table}

We use the same dataset as \cite{Benjamin2003a}, \cite{Chernozhukov2004} and
\cite{Wuthrich2019}. The data consists of a sample of 9,910 households from
the 1991 SIPP\footnote{The original data have 9,915 households, but we follow
\cite{Benjamin2003a} and delete the five observations with zero or negative
income. Descriptive statistics are available in Table 1 in
\cite{Benjamin2003a} and in Tables 1 and 2 in \cite{Chernozhukov2004}.}. The
outcomes of interest are net financial assets, and total wealth. For the
(instrument) propensity score estimation, we adopt a logistic specification,
and use all two-way interactions between income, log-income, age, family size,
years of education, dummies for homeownership, marital status, two-earner
status, defined benefit pension status, and individual retirement account
participation status. To assess the reliability of this parametric PS model,
we apply the specification test of \cite{SantAnna2018} with 1,000 bootstrap
draws, and fail to reject the null of the propensity score model being
correctly specified at the 10\% level.

Panel A (Panel B) of Table \ref{tab:asset1} shows the point estimates and
standard errors (in parentheses) for the effect of 401(k) eligibility
(participation among compliers) on net financial assets and total wealth. We
present IPW estimators for the ATE, QTE$\left(  0.25\right)  $, QTE$\left(
0.5\right)  $ and QTE$\left(  0.75\right)  $, and for the LATE, LQTE$\left(
0.25\right)  $, LQTE$\left(  0.5\right)  $ and LQTE$\left(  0.75\right)  $
using the same PS estimation methods as in the simulation exercise in Section
\ref{simu}, except the IPS and LIPS estimators based on the indicator
weighting function, as they tend to be numerically unstable when the dimension of covariates is moderate.

\begin{figure}[H]
\begin{center}
\begin{adjustbox}{scale=0.8 }
\includegraphics{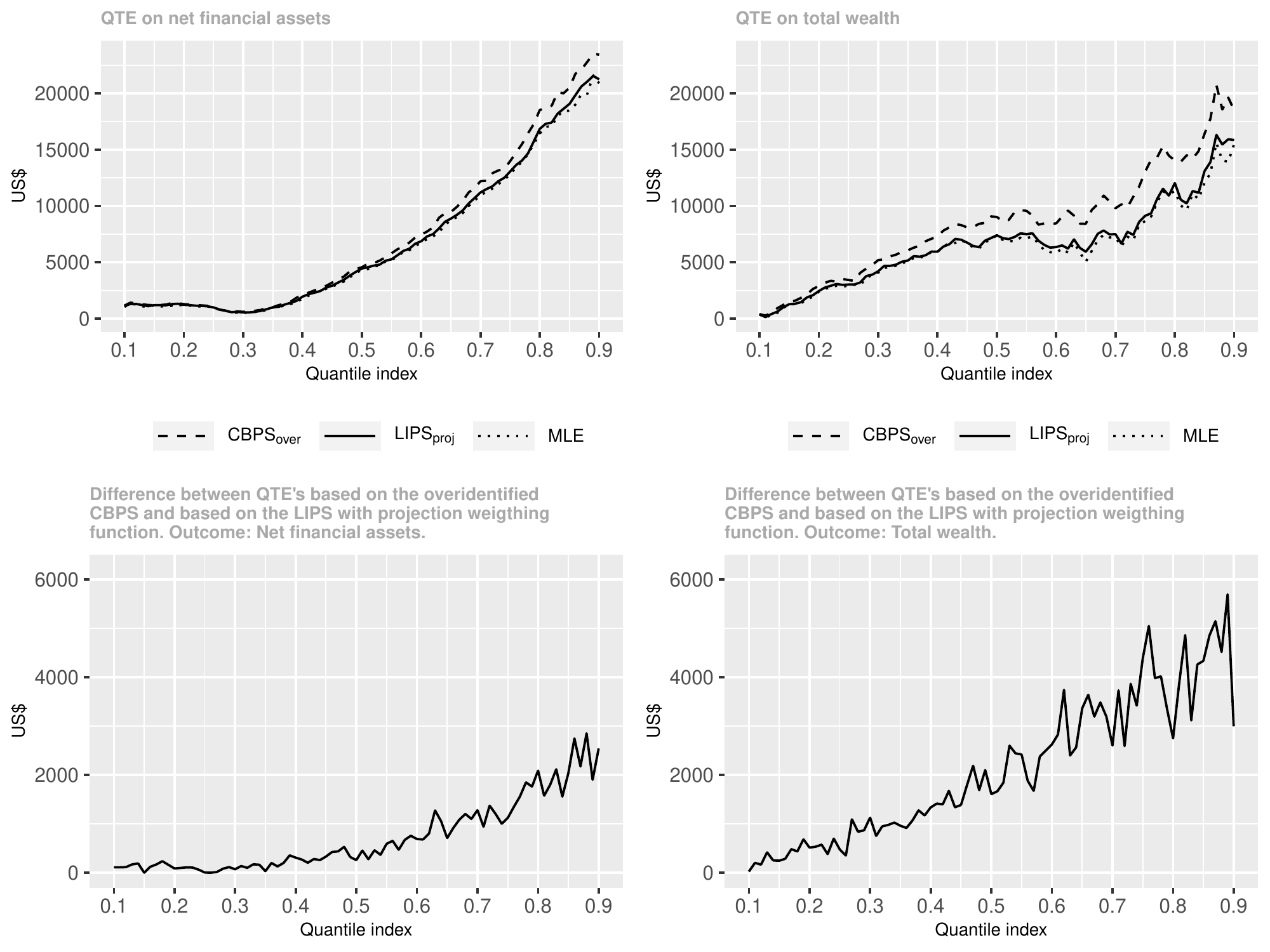}
\end{adjustbox}
\end{center}
\par
\vspace*{-5mm}\caption{{Estimated quantile treatment effects of 401(k)
eligibility on different wealth measures.}}%
\label{fig.qte}%
\end{figure}

The results in Panel A suggest that 401(k) eligibility has a positive and
significant average impact on both net financial assets and total wealth and
that the effect is more pronounced at the higher quantiles. When one compares
the treatment effect measures across different PS estimation methods, we see
that the results tend to be similar for net financial assets; for total
wealth, we note that estimators based on the overidentified CBPS estimator
suggest much larger effects of 401(k) eligibility at higher quantiles than
those based on our proposed IPS estimators; see Figure \ref{fig.qte} for a
more detailed comparison between the QTE estimates based on the IPS with
projection weighting function, overidentified CBPS (the default in the
\texttt{CBPS R} package), and those based on ML.

\begin{figure}[pth]
\begin{center}
\begin{adjustbox}{scale=0.8 }
\includegraphics{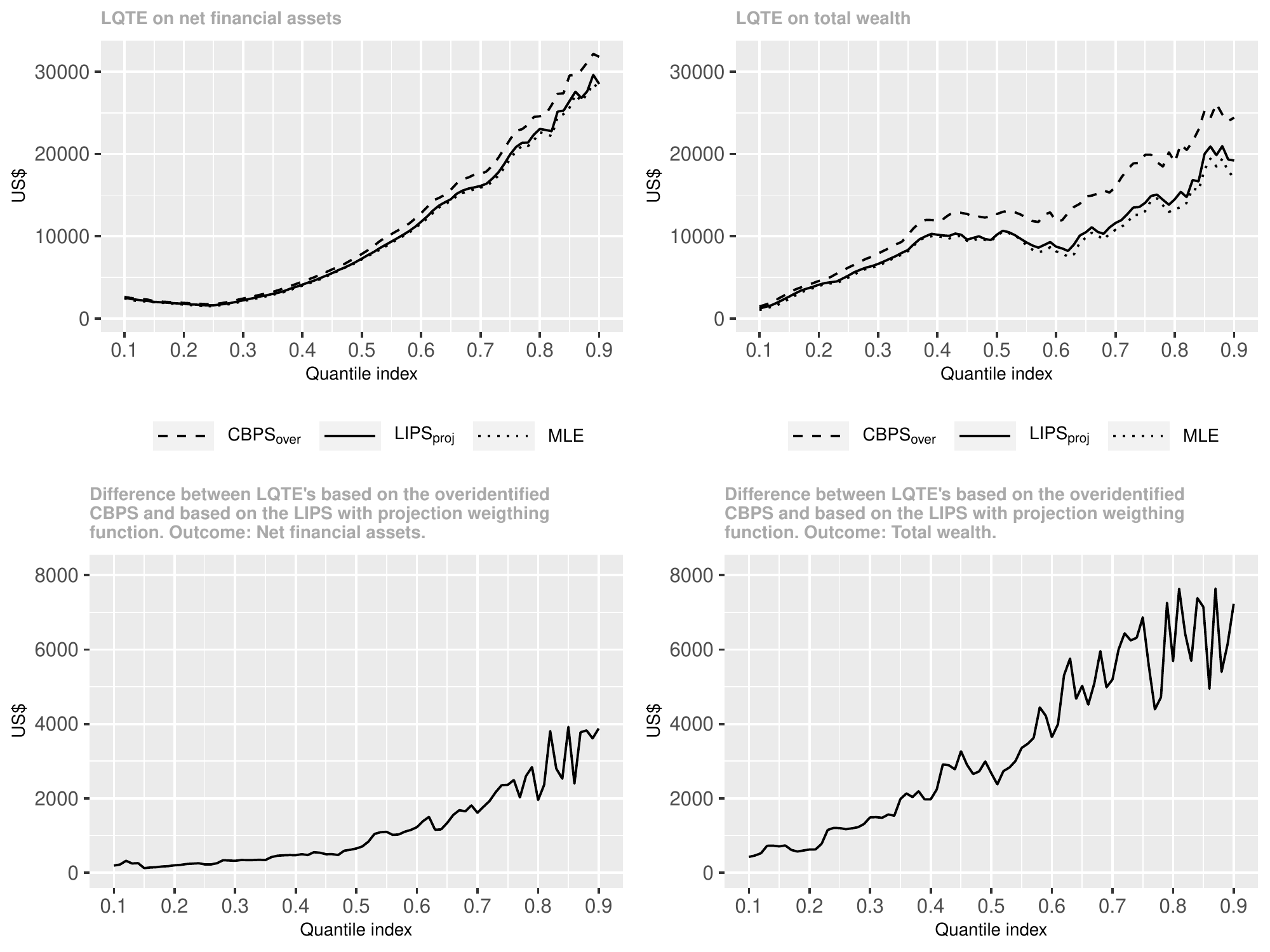}
\end{adjustbox}
\end{center}
\par
\vspace*{-5mm}\caption{{Estimated local quantile treatment effects of 401(k)
participation on different wealth measures.}}%
\label{fig.lqte}%
\end{figure}

The results in Panel B paint a similar picture as those in Panel A: 401(k)
participation tends to have a positive and significant average impact on both
measures of wealth, and the effect is more pronounced at the right tail of the
wealth measures. As we illustrated in Figure \ref{fig.lqte}, there are
quantitative differences between the LQTE estimates based on different PS
estimation methods, with those based on the overidentified instrument CBPS
suggesting much larger effects than the other estimation methods, though the
shape of the LQTE function is similar across specifications.

\section{Conclusion \label{sec:conc}}

In this article, we proposed a framework to estimate propensity score
parameters such that, instead of targeting to balance only some specific
moments of covariates, it aims to balance \textit{all} functions of
covariates. The proposed estimator is of the minimum distance type, and is
data-driven, $\sqrt{n}$-consistent, asymptotically normal, and admits an
asymptotic linear representation that facilitates the study of inverse
probability weighted estimators in a unified manner. Importantly, we have
shown that our framework can accommodate the empirically relevant situation
under which treatment allocation is endogenous. We derived the large sample
properties of average, distributional and quantile treatment effect estimator
based on the proposed integrated propensity scores, and illustrated its
attractive properties via a Monte Carlo study and two empirical applications.

Although this paper devoted most of its attention to forming IPW-type
treatment effect estimators, we note that sometimes researchers are willing to
consider an outcome regression model, on top of the propensity score model. In
such cases, we stress that one can easily combine our IPS estimation procedure
with such outcome regression model to form doubly-robust, locally efficient
treatment effect estimators, see, e.g., \cite{Sloczynski2018} and references
therein. Perhaps even better, one can use the integrated moment approach
adopted in this paper to estimate not only the propensity score, but also the
outcome regression model. We leave the detailed discussion of such procedure
for future research.

\onehalfspacing{\small
\bibliographystyle{jasa}
\bibliography{IPS}
}

\end{document}